\documentclass[unsortedaddress,aps,preprint]{revtex4}
\usepackage{graphicx}
\usepackage{dcolumn}
\usepackage{amsmath}
\usepackage{amssymb}
\usepackage{amsfonts}
\usepackage{bm}
\usepackage{color}
\usepackage[normalem]{ulem}
\usepackage{hyperref}
\hypersetup{colorlinks=true,linkcolor=blue,urlcolor=magenta}
\newcommand{\be}{\begin{equation}}
\newcommand{\ee}{\end{equation}}
\newcommand{\ba}{\begin{eqnarray}}
\newcommand{\ea}{\end{eqnarray}}

\begin{document}
\title{Entropy Multiparticle Correlation Expansion for a Crystal}
\author{Santi Prestipino$^1$\footnote{Email: {\tt sprestipino@unime.it}} and Paolo V. Giaquinta$^1$}
\affiliation{$^1$Universit\`a degli Studi di Messina,\\Dipartimento di Scienze Matematiche ed Informatiche, Scienze Fisiche e Scienze della Terra,\\Viale F. Stagno d'Alcontres 31, 98166 Messina, Italy}

\begin{abstract}As first shown by H.~S.~Green in 1952, the entropy of a classical fluid of identical particles can be written as a sum of many-particle contributions, each of them being a distinctive functional of all spatial distribution functions up to a given order. By revisiting the combinatorial derivation of the entropy formula, we argue that a similar correlation expansion holds for the entropy of a crystalline system. We discuss how one- and two-body entropies scale with the size of the crystal, and provide fresh numerical data to check the expectation, grounded on theoretical arguments, that both entropies are extensive quantities.
\end{abstract}
\maketitle
\section{Introduction}

The entropy multiparticle correlation expansion (MPCE) is an elegant statistical-mechanical formula that entails the possibility to reconstruct the total entropy of a many-particle system term by term, including at each step of summation the integrated contribution from spatial correlations between a specified number of particles.

The original derivation of the entropy MPCE is found in a book by H. S. Green (1952)~\cite{Green}. Green's expansion applies for the canonical ensemble (CE). In 1958, Nettleton and M. S. Green derived an apparently different expansion valid in the grand-canonical ensemble (GCE)~\cite{Nettleton}. It took the ingenuity of Baranyai and Evans to realize, in 1989, that the CE expansion can indeed be reshuffled in such a way as to become formally equivalent to the GCE expansion~\cite{Baranyai}.

A decisive step forward was eventually taken by Schlijper~\cite{Schlijper} and An~\cite{An}, who have highlighted the similarity of the entropy formula to a cumulant expansion, as well as the close relationship with the cluster variation method (see, e.g., Ref.~\cite{Pelizzola}). Other papers where in various ways it is emphasized the combinatorial content of the entropy MPCE are Refs.~\cite{Hernando,Prestipino1,Prestipino2,D'Alessandro}.

Since the very beginning it has been clear that the successive terms in the entropy expansion for a homogeneous fluid are not all of equal importance. In particular, the contributions from correlations between more than two particles are only sizable at moderate and higher densities. However, while the two-body entropy is easily accessed in a simulation, computing the higher-order entropy terms is a prohibitive task (see, however, Ref.~\cite{Maffioli}). Hence, the only viable method to compute the total entropy in a simulation remains thermodynamic integration (see e.g. \cite{Abramo}). The practical interest for the entropy expansion has thus shifted towards the {\em residual multiparticle entropy} (RMPE), defined as the difference between excess entropy and two-body entropy. The RMPE is a measure of the impact of non-pair multiparticle correlations on the entropy of the fluid. For hard spheres, Giaquinta and Giunta have observed that the RMPE changes sign from negative to positive very close to freezing~\cite{Giaquinta1}. At low densities the RMPE is negative, reflecting a global reduction (largely driven by two-body correlations) of the phase space available to the system as compared to the ideal gas. The change of sign of the RMPE close to freezing indicates that fluid particles, which at high enough densities are forced by more stringent packing constraints, start exploring, now in a cooperative way, a different structural condition on a local scale, preluding to crystallization on a global scale. Since the original observation in \cite{Giaquinta1}, a clear correspondence between the RMPE zero and the ultimate threshold for spatial homogeneity in the system has been found in many simple and complex fluids~\cite{Giaquinta2,Saija1,Donato,Saija2,Costa,Prestipino3,Saija3,Speranza,Prestipino4,Banerjee,Santos}, thus leading to the belief that the vanishing of the RMPE is a signature of an impending structural or thermodynamic transition of the system from a less ordered to a more spatially organized condition (freezing is just an example of many). Albeit empirical, this {\em entropic criterion} is a valid alternative to the far more demanding exact free-energy methods when a rough estimate of the transition point is deemed sufficient. For a simple discussion of the interplay between entropy and ordering, the reader is referred to Ref.~\cite{Frenkel}; see instead Refs.~\cite{Speedy,Berthier} for general considerations about the entropy of disordered solids.

A pertinent question to ask is what happens to the RMPE on the solid side of the phase boundary, considering that an entropy expansion also holds for the crystal. This is precisely the problem addressed in this paper. Can the scope of the entropic criterion be extended in such a way that it also applies for melting? As it turns out, we can offer no definite answer to this question, since theory alone does not go too far and we run into a serious computational bottleneck: while the formulae are clear and the numerical procedure is straightforward, it is extremely hard to obtain reliable data for the two-body entropy of a three-dimensional crystal. We have only carried out a limited test on a triangular crystal of hard disks, but our results are affected by finite-size artifacts that make them inconclusive. Nevertheless, a few firm points have been established: 1) The approximate entropy expressions obtained by truncating the MPCE at a given order can all be derived from an explicit functional of the correlation functions up to that order; 2) The one-body entropy for a crystal is an extensive quantity (the same is held to be true for the two-body entropy, but our arguments are not sufficient for a proof); 3) The peaks present in the crystal one-body density have a nearly Gaussian shape; 4) We have also clarified the role of lattice symmetries in dictating the structure of the two-body density, which is explicitly determined at zero temperature.

This paper is organized as follows. In Section 2 we resume the formalism of the entropy expansion for homogeneous fluids and provide the basic tools needed for its extension to crystals. Then, in Section 3 we exploit the symmetries of one- and two-body density functions to predict the scaling of one- and two-body entropies with the size of the crystal. The final Section 4 is reserved to concluding remarks.

\section{Derivation of the entropy MPCE}

In this Section, we collect a number of well-established results on the entropy MPCE, with the only purpose of setting the language and notation for the rest of the paper. First, we recall the derivation of the entropy formula for a one-component system of classical particles in the canonical ensemble. Such an ensemble choice is by no means restrictive since, as we show next, it is always possible to take advantage of the sum rules obeyed by the canonical correlation functions to arrange the entropy MPCE in an ensemble-invariant form. Then, in the following Section we present an application of the formalism to crystals.

The canonical partition function of a system of $N$ classical particles of mass $m$ at temperature $T$ is $Z_N=Z_N^{\rm id}Z_N^{\rm exc}$, where the ideal and excess parts are given by
\be
Z_N^{\rm id}=\frac{1}{N!}\left(\frac{V}{\Lambda^3}\right)^N
\,\,\,\,\,\,{\rm and}\,\,\,\,\,\,
Z_N^{\rm exc}=\frac{1}{V^N}\int{\rm d}^3R_1\cdots{\rm d}^3R_N\,e^{-\beta U({\bf R}^N)}\,.
\label{eq-1}
\ee
In Eq.~(\ref{eq-1}), $V$ is the system volume, $\beta=1/(k_BT)$, $\Lambda=h/\sqrt{2\pi mk_BT}$ is the thermal wavelength, and $U({\bf R}^N)$ is an arbitrary potential energy. As the particles are identical, for each $n=1,2,\ldots,N$ the cumulative sum of all $n$-body terms in $U$ is invariant under permutations of particle coordinates (we can also say that $U$ is $S_N$-invariant, $S_N$ being the symmetric group of the permutations on $N$ symbols). The CE average of a function $f$ of coordinates reads
\be
\left\langle f({\bf R}^N)\right\rangle\equiv\frac{1}{V^N}\int{\rm d}^3R_1\cdots{\rm d}^3R_N\,f({\bf R}^N)\pi_{\rm can}({\bf R}^N)
\,\,\,\,\,\,{\rm with}\,\,\,\,\,\,
\pi_{\rm can}({\bf R}^N)=\frac{e^{-\beta U({\bf R}^N)}}{Z_N^{\rm exc}}\,,
\label{eq-2}
\ee
where $\pi_{\rm can}({\bf R}^N)$ is the configurational part of the canonical density function. Finally, the excess entropy $S_N^{\rm exc}\equiv S_N-S_N^{\rm id}$ reads
\be
\frac{S_N^{\rm exc}}{k_B}=-\frac{1}{V^N}\int{\rm d}^3R_1\cdots{\rm d}^3R_N\,\pi_{\rm can}({\bf R}^N)\ln\pi_{\rm can}({\bf R}^N)=-\left\langle\ln\pi_{\rm can}({\bf R}^N)\right\rangle\,.
\label{eq-3}
\ee

We define a set of {\em marginal density functions} (MDF) by
\ba
P^{(N)}({\bf R}^N)&=&\pi_{\rm can}({\bf R}^N)\,;
\nonumber \\
P^{(n)}({\bf R}^n)&=&\frac{1}{V^{N-n}}\int{\rm d}^3R_{n+1}\cdots{\rm d}^3R_N\,\pi_{\rm can}({\bf R}^N)\,\,\,\,\,\,(n=1,\ldots,N-1)\,.
\label{eq-4}
\ea
Owing to $S_N$-invariance of $\pi_{\rm can}$, it makes no difference which vector radii are integrated out in Eq.\,(\ref{eq-4}); hence, $P^{(n)}({\bf r}^n)$ is $S_n$-invariant (for example, $P^{(2)}({\bf r},{\bf r}')=P^{(2)}({\bf r}',{\bf r})$). The following properties are obvious:
\be
\frac{1}{V^n}\int{\rm d}^3R_{1}\cdots{\rm d}^3R_n\,P^{(n)}({\bf R}^n)=1
\,\,\,\,\,\,{\rm and}\,\,\,\,\,\,
\frac{1}{V}\int{\rm d}^3R_{n+1}\,P^{(n+1)}({\bf R}^{n+1})=P^{(n)}({\bf R}^n)\,,
\label{eq-5}
\ee
Then, the {\em $n$-body density functions} (DF), for $n=1,\ldots,N$, can be expressed as
\be
\rho^{(n)}({\bf r}^n)\equiv\left\langle\sideset{}{^\prime}\sum_{i_1\ldots i_n}\delta^3({\bf R}_{i_1}-{\bf r}_1)\cdots\delta^3({\bf R}_{i_n}-{\bf r}_n)\right\rangle=\frac{N!}{(N-n)!}\frac{P^{(n)}({\bf r}^n)}{V^n}\,,
\label{eq-6}
\ee
where the sum in (\ref{eq-6}) is carried out over all $n$-tuples of distinct particles (for example, the sum for $n=2$ contains $N(N-1)$ terms). We note that $P^{(1)}=1$ and $\rho^{(1)}=N/V\equiv\rho$ if no one-body term is present in $U$, i.e., if no external potential acts on the particles (then $U$ is translationally invariant). $P^{(1)}({\bf r})/V$ is the probability density of finding a particle in ${\bf r}$; hence, $\rho^{(1)}({\bf r})=NP^{(1)}({\bf r})/V$ is the number density at ${\bf r}$. Similarly, $P^{(2)}({\bf r},{\bf r}')/V^2$ is the probability density of finding one particle in ${\bf r}$ and another particle in ${\bf r}'$, hence $\rho^{(2)}({\bf r},{\bf r}')=N(N-1)P^{(2)}({\bf r},{\bf r}')/V^2$ is the density of the number of particle pairs at $({\bf r},{\bf r}')$. As ${\bf r}'$ increasingly departs from ${\bf r}$, the positions of two particles become less and less correlated, until $P^{(2)}({\bf r},{\bf r}')=P^{(1)}({\bf r})P^{(1)}({\bf r}')$ at infinite distance. We stress that this {\em cluster property} holds in full generality, even for a broken-symmetry phase.

The {\em $n$-body reduced density functions}, for $n=2,\ldots,N$, read
\ba
&&g^{(n)}({\bf r}^n)\equiv\frac{\rho^{(n)}({\bf r}^n)}{\rho^{(1)}({\bf r}_1)\cdots \rho^{(1)}({\bf r}_n)}=\left(1-\frac{1}{N}\right)\cdots\left(1-\frac{n-1}{N}\right)Q^{(n)}({\bf r}^n)
\nonumber \\
&&{\rm with}\,\,\,\,\,\,
Q^{(n)}({\bf r}^n)=\frac{P^{(n)}({\bf r}^n)}{P^{(1)}({\bf r}_1)\cdots P^{(1)}({\bf r}_n)}\,.
\label{eq-7}
\ea
These functions fulfil the property
\be
\frac{1}{V}\int{\rm d}^3R_{n+1}\,P^{(1)}({\bf R}_{n+1})g^{(n+1)}({\bf R}^{n+1})=\left(1-\frac{n}{N}\right)g^{(n)}({\bf R}^n)\,,
\label{eq-8}
\ee
which also holds for $n=1$ if we define $g^{(1)}\equiv 1$. For a homogeneous fluid, $g^{(2)}({\bf r},{\bf r}')=g(|{\bf r}-{\bf r}'|)$. From now on, we adopt the shorthand notations $P_{12\ldots n}=P^{(n)}({\bf R}^n)$ and $Q_{12\ldots n}=Q^{(n)}({\bf R}^n)$. Moreover, any integral of the kind $V^{-n}\int{\rm d}^3R_{1}\cdots{\rm d}^3R_n\,(\cdots)$ is hereafter denoted as $\int(\cdots)$. For example, Eqs.~(\ref{eq-3}) and (\ref{eq-4}) indicate that $S_N^{\rm exc}/k_B=-\int P_{12\ldots N}\ln P_{12\ldots N}$.

To build up the CE expansion term by term, our strategy is to consider a progressively larger number of particles. For a one-particle system, the excess entropy in units of the Boltzmann constant is $S_1^{\rm exc}/k_B=-\int P_1\ln P_1$, leading to a first-order approximation to the excess entropy of a $N$-particle system in the form $S_N^{\rm exc}/k_B\approx S_N^{(1)}/k_B\equiv -N\int P_1\ln P_1$ (that is, each particle contributes to the entropy independently of the other particles). For a two-particle system, the excess entropy is $S_2^{(1)}$ plus a remainder $k_BR_2$, given by:
\be
R_2\equiv\frac{S_2^{\rm exc}-S_2^{(1)}}{k_B}=-\int P_{12}\ln P_{12}+2\int P_1\ln P_1=-\int P_{12}\ln Q_{12}\,.
\label{eq-9}
\ee
Equation (\ref{eq-9}) suggests a second-order approximation for $S_N^{\rm exc}$, where each distinct pair of particles contributes the same two-body residual term to the entropy:
\be
\frac{S_N^{(2)}}{k_B}=-N\int P_1\ln P_1-{N\choose 2}\int P_{12}\ln Q_{12}\,.
\label{eq-10}
\ee
Notice that Eq.~(\ref{eq-10}) is exact for $N=2$, i.e., $S_2^{(2)}=S_2^{\rm exc}$. Similarly, for a three-particle system the excess entropy is $S_3^{(2)}$ plus a remainder $k_BR_3$:
\ba
R_3\equiv\frac{S_3^{\rm exc}-S_3^{(2)}}{k_B}&=&-\int P_{123}\ln P_{123}+3\int P_1\ln P_1+{3\choose 2}\int P_{12}\ln Q_{12}
\nonumber \\
&=&-\int P_{123}\ln Q_{123}+{3\choose 2}\int P_{12}\ln Q_{12}\,.
\label{eq-11}
\ea
Whence, a third-order approximation follows for $S_N^{\rm exc}$ in the form
\be
\frac{S_N^{(3)}}{k_B}=-N\int P_1\ln P_1-{N\choose 2}\int P_{12}\ln Q_{12}-{N\choose 3}\left[\int P_{123}\ln Q_{123}-{3\choose 2}\int P_{12}\ln Q_{12}\right]\,.
\label{eq-12}
\ee
Again, $S_3^{(3)}=S_3^{\rm exc}$. Equation (\ref{eq-12}) reproduces the first three terms in the rhs of equation (5.9) of Ref.~\cite{Prestipino1}, and one may legitimately expect that the further terms in the entropy expansion are similarly obtained by arguing for $N=4,5,\ldots$ like we did for $N=1,2,3$ (see the proof in \cite{Prestipino1}).

The general entropy formula finally reads:
\ba
\frac{S_N^{\rm exc}}{k_B}&=&-\int P_{12\ldots N}\ln P_{12\ldots N}=-N\int P_1\ln P_1-\int P_{12\ldots N}\ln Q_{12\ldots N}
\nonumber \\
&=&-N\int P_1\ln P_1-\sum_{n=2}^N{N\choose n}\sum_{a=2}^n(-1)^{n-a}{n\choose a}\int P_{1\ldots a}\ln Q_{1\ldots a}\,.
\label{eq-13}
\ea
This equation is trivially correct since, for any finite sequence $\{c_a\}$ of numbers,
\be
c_N=\sum_{n=2}^N{N\choose n}\sum_{a=2}^n(-1)^{n-a}{n\choose a}c_a\,.
\label{eq-14}
\ee
To prove (\ref{eq-14}) it is sufficient to observe that, for each fixed $k=2,\ldots,N$, the coefficient of $c_k$ in the above sum is
\ba
\sum_{n=k}^N(-1)^{n-k}{N\choose n}{n\choose k}&=&\sum_{n=0}^{N-k}(-1)^n{N\choose n+k}{n+k\choose k}={N\choose k}\sum_{n=0}^{N-k}(-1)^n{N-k\choose n}
\nonumber \\
&=&\left\{
\begin{array}{cl}
0, & \,\,\,{\rm for}\,\,2\le k<N\\
1, & \,\,\,{\rm for}\,\,k=N\,.
\end{array}
\right.
\label{eq-15}
\ea
A more compact entropy formula is
\be
\frac{S_N^{\rm exc}}{k_B}=-\sum_{n=1}^N{N\choose n}\sum_{a=1}^n(-1)^{n-a}{n\choose a}\int P_{1\ldots a}\ln P_{1\ldots a}\,,
\label{eq-16}
\ee
which follows from
\be
c_N=\sum_{n=1}^N{N\choose n}\sum_{a=1}^n(-1)^{n-a}{n\choose a}c_a\,.
\label{eq-17}
\ee

The entropy expansion, (\ref{eq-13}) or (\ref{eq-16}), is only valid in the CE. Eliminating $Q_{1\ldots a}$ in favor of $g_{1\ldots a}$ by Eq.~(\ref{eq-7}), an overall constant comes out of the integral in Eq.~(\ref{eq-13}), namely
\be
\sum_{n=2}^N{N\choose n}\sum_{a=2}^n(-1)^{n-a}{n\choose a}\ln\frac{(N-1)\cdots(N-a+1)}{N^{a-1}}\,,
\label{eq-18}
\ee
which, by Eq.~(\ref{eq-15}), equals $\ln\left(N!/N^N\right)$; this term exactly cancels an identical term present in the ideal-gas entropy. In the end, a modified entropy MPCE emerges:
\ba
\frac{S_N}{k_B}=N\left[\frac{3}{2}-\ln(\rho\Lambda^3)\right]-N\int P_1\ln P_1-\sum_{n=2}^N{N\choose n}\sum_{a=2}^n(-1)^{n-a}{n\choose a}\int P_{1\ldots a}\ln g_{1\ldots a}\,.
\label{eq-19}
\ea
Notice that the first term in the rhs differs by $N$ from the ideal-gas entropy expression in the thermodynamic limit. In order that Eq.~(\ref{eq-19}) conforms to the GCE entropy expansion, for each $n$ a suitable {\em fluctuation integral} of value $-N/[n(n-1)]$ should be summed to (and subtracted from) the $n$-th term in the expansion. For example, using Eqs.~(\ref{eq-7}) and (\ref{eq-8}) the second-order term in (\ref{eq-13}) can be rewritten as
\small
\ba
-{N\choose 2}\int\frac{{\rm d}^3r_1{\rm d}^3r_2}{V^2}P_{12}\ln\frac{P_{12}}{P_1P_2}&=&{N\choose 2}\ln\frac{N-1}{N}-{N\choose 2}\int\frac{{\rm d}^3r_1{\rm d}^3r_2}{V^2}P_1P_2g_{12}\ln g_{12}
\nonumber \\
&=&{N\choose 2}\ln\frac{N-1}{N}+\frac{N}{2}-\frac{1}{2}\rho^2\int{\rm d}^3r_1{\rm d}^3r_2\,P_1P_2\left(g_{12}\ln g_{12}-g_{12}+1\right)\,.
\label{eq-20}
\ea
\normalsize
Overall, the extra constants appearing in each term of the entropy formula (for example, the quantity $N/2$ in Eq.~(\ref{eq-20})) add to $N$. By absorbing such a $N$ in the first term of (\ref{eq-19}) we recover the ideal-gas entropy in the thermodynamic limit, and the CE expansion becomes formally identical to the grand-canonical MPCE~\cite{Prestipino2}.

In Appendix A we present another derivation of the entropy formula in the CE, which is closer in spirit to the one given by H. S. Green. In parallel, we show that the approximation obtained by truncating the MPCE at a given order can be derived from a modified $P_{12\ldots N}$ distribution, which is an explicit functional of the spatial correlation functions up to that order.

\section{The first few terms in the expansion of crystal entropy}
The entropy expansion in the CE is formally identical for a fluid system and a crystal, since the origin of (\ref{eq-13}) is purely combinatorial. However, the DFs of the two phases are radically different: most notably, while $P_1=1$ and $\rho^{(1)}=\rho$ for a homogeneous fluid, the one-body density is spatially structured for a crystal --- at least once the degeneracy due to translations and point-group operations has been lifted; we stress that $P_1\neq 1$ only provided that a specific determination of the crystal is taken, since otherwise $P_1=1$ also in the ``delocalized'' crystalline phase. In practice, in order to fix a crystal in space we should imagine to apply a suitable symmetry-breaking external potential, whose strength is sent to zero after statistical averages have been carried out (in line with Bogoliubov's advice to interpret statistical averages of broken-symmetry phases as {\em quasiaverages}~\cite{Baus}, which amounts to send the strength of the symmetry-breaking potential to zero only {\em after} the thermodynamic limit has been taken). A way to accomplish this task is to constrain the position of just one particle. When periodic boundary conditions are applied, holding one particle fixed will be enough to break the continuous symmetries of free space. As $N$ grows, the effect of the external potential becomes weaker and weaker since it does not scale with the size of the system.

\subsection{One-body entropy}

A reasonable form of one-body density for a three-dimensional Bravais crystal without defects is the Tarazona ansatz~\cite{Tarazona}:
\be
\rho^{(1)}({\bf r})=\left(\frac{\alpha}{\pi}\right)^{3/2}\sum_{\bf R}e^{-\alpha({\bf r}-{\bf R})^2}=\rho\sum_{\bf G}e^{-\frac{G^2}{4\alpha}}e^{i{\bf G}\cdot{\bf r}}\,,
\label{eq-21}
\ee
where $\alpha>0$ is a temperature-dependent parameter, the {\bf R}'s are direct-lattice vectors, and the {\bf G}'s are reciprocal-lattice vectors [recall that ${\bf G}\cdot{\bf R}=2\pi m$ with $m\in\mathbb{Z}$ and $\int_V{\rm d}^3r\,\exp\{i({\bf G}+{\bf G}')\cdot{\bf r}\}=V\delta_{{\bf G}',-{\bf G}}$]. Equation (\ref{eq-21}) is a rather generic form of crystal density, which recently we have also applied in a different context~\cite{Prestipino5}. More generally, the one-body density appropriate to a perfect crystal must obey $\rho^{(1)}({\bf r}+{\bf R})=\rho^{(1)}({\bf r})$ for all ${\bf R}$, and is thus necessarily of the form
\be
\rho^{(1)}({\bf r})=\sum_{\bf G}\widetilde{u}_{\bf G}e^{i{\bf G}\cdot{\bf r}}\,\,\,\,\,\,{\rm with}\,\,\,\,\,\,\widetilde{u}_{\bf G}^*=\widetilde{u}_{-{\bf G}}\,.
\label{eq-22}
\ee
Since $\int_V{\rm d}^3r\,\rho^{(1)}({\bf r})=N$, it soon follows $\widetilde{u}_0=\rho$. Calling ${\cal C}$ a primitive cell and $v_0$ its volume, $\widetilde{u}_{\bf G}=v_0^{-1}\int_{\cal C}{\rm d}^3r\,\rho^{(1)}({\bf r})\exp\{-i{\bf G}\cdot{\bf r}\}\rightarrow 0$ as $G\rightarrow\infty$ (by the Riemann-Lebesgue lemma). In real space, a legitimate $\rho^{(1)}({\bf r})$ function is $\sum_{\bf R}\phi({\bf r}-{\bf R})$ with $\int{\rm d}^3r\,\phi({\bf r})=1$ (integration bounds are left unspecified when the integral is over a macroscopic $V$). In the zero-temperature/infinite-density limit, particles sit at the lattice sites and the one-body density then becomes
\be
\rho^{(1)}({\bf r})=\sum_{\bf R}\delta^3({\bf r}-{\bf R})\,.
\label{eq-23}
\ee
Equation (\ref{eq-23}) is also recovered from Eq.~(\ref{eq-21}) in the $\alpha\rightarrow\infty$ limit.

For a crystalline solid, the {\em one-body entropy}, that is the first term in the expansion of excess entropy, is (in units of $k_B$):
\be
S_1\equiv-\frac{N}{V}\int{\rm d}^3r_1\,P_1\ln P_1=-\int{\rm d}^3r_1\,\rho^{(1)}({\bf r}_1)\ln\frac{\rho^{(1)}({\bf r}_1)}{\rho}\,.
\label{eq-24}
\ee
One may wonder whether the integral in (\ref{eq-24}) is ${\cal O}(N)$ in the infinite-size limit. The answer is affirmative, and a simple argument goes as follows. Let $\rho^{(1)}({\bf r})$ be $\sum_{\bf R}\phi({\bf r}-{\bf R})$; if $\phi({\bf r})$ is strongly localized near ${\bf r}=0$, then $\rho^{(1)}({\bf r})\simeq\phi({\bf r})$ in the cell around ${\bf R}=0$ and $S_1\simeq -N\int_{\cal C}{\rm d}^3r\,\phi({\bf r})\ln\left\{\phi({\bf r})/\rho\right\}={\cal O}(N)$. Actually, we can provide a rigorous proof that $S_1$ is negative-semidefinite and its absolute value does not grow faster than $N$. Using $\ln x\le x-1$ for $x>0$ and $x\ln x\ge x-1$ for any $x\ge 0$, we obtain
\be
0=\rho\int{\rm d}^3r_1\left(\frac{\rho_1}{\rho}-1\right)\le\int{\rm d}^3r_1\,\rho_1\ln\frac{\rho_1}{\rho}\le\int{\rm d}^3r_1\,\rho_1\left(\frac{\rho_1}{\rho}-1\right)=\rho^{-1}\int{\rm d}^3r_1\,\rho_1^2-N\,.
\label{eq-25}
\ee
To estimate $\int{\rm d}^3r_1\,\rho_1^2$ we employ the one-body density in (\ref{eq-21}), which is sufficiently generic for our purposes:
\be
\rho^{-1}\int{\rm d}^3r\left(\rho^{(1)}({\bf r})\right)^2=\rho\sum_{{\bf G},{\bf G}'}e^{-\left(\frac{G^2}{4\alpha}+\frac{G^{\prime 2}}{4\alpha}\right)}\int{\rm d}^3r\,e^{i({\bf G}+{\bf G}')\cdot{\bf r}}=N\sum_{\bf G}e^{-\frac{G^2}{2\alpha}}
\label{eq-26}
\ee
(the above result is nothing but Parseval's theorem as applied to (\ref{eq-21})). The sum in the rhs of Eq.~(\ref{eq-26}) is the three-dimensional analog of a Jacobi theta function (see, e.g., \cite{Prestipino6}), whose value is ${\cal O}(1)$ for $\alpha>0$. Therefore, it follows from Eqs.~(\ref{eq-25}) and (\ref{eq-26}) that the one-body entropy is at most ${\cal O}(N)$.

\subsection{Two-body entropy}
We now move to the problem of evaluating the {\em two-body entropy} $S_2$ for a crystal. For a homogeneous fluid, $S_2$ is an extensive quantity which, in $k_B$ units, is equal to
\be
{\rm fluid}:\qquad S_2=-2\pi\rho N\int_0^\infty{\rm d}r\,r^2\left(g(r)\ln g(r)-g(r)+1\right)\,.
\label{eq-27}
\ee
For a crystal, we have from Eq.~(\ref{eq-20}) that
\be
S_2=-\frac{1}{2}\rho^2\int{\rm d}^3r_1{\rm d}^3r_2\,P_1P_2\left(g_{12}\ln g_{12}-g_{12}+1\right)\,.
\label{eq-28}
\ee
As $x\ln x\ge x-1$ for $x>0$, $S_2$ is usually negative and zero exclusively for $g_{12}=1$. In terms of density functions, $S_2$ is written as
\be
S_2=-\frac{1}{2}\int{\rm d}^3r_1{\rm d}^3r_2\left(\rho^{(2)}({\bf r}_1,{\bf r}_2)\ln\frac{\rho^{(2)}({\bf r}_1,{\bf r}_2)}{\rho^{(1)}({\bf r}_1)\rho^{(1)}({\bf r}_2)}-\rho^{(2)}({\bf r}_1,{\bf r}_2)+\rho^{(1)}({\bf r}_1)\rho^{(1)}({\bf r}_2)\right)\,,
\label{eq-29}
\ee
We show below that Eq.~(\ref{eq-29}) can be expressed as a radial integral, i.e., in a way similar to the two-body entropy for a fluid.

We can assign a radial structure to crystals by appealing to a couple of functions introduced in \cite{Rascon}, namely
\be
\rho^2\widetilde{g}(r)=\int\frac{{\rm d}^3r_1}{V}\int\frac{{\rm d}^2\Omega}{4\pi}\,\rho^{(2)}({\bf r}_1,{\bf r}_1+{\bf r})
\label{eq-30}
\ee
and
\be
\rho^2\widetilde{g}_0(r)=\int\frac{{\rm d}^3r_1}{V}\int\frac{{\rm d}^2\Omega}{4\pi}\,\rho^{(1)}({\bf r}_1)\rho^{(1)}({\bf r}_1+{\bf r})\,,
\label{eq-31}
\ee
where the inner integrals are over the direction of {\bf r}. For a homogeneous fluid, $\widetilde{g}(r)=g(r)$ and $\widetilde{g}_0(r)=1$. The Authors of Ref.~\cite{Rascon} have sketched the profile of $\widetilde{g}(r)$ and $\widetilde{g}_0(r)$ for a crystal; both functions show narrow peaks at neighbor positions in the lattice, with an extra peak at zero distance for $\widetilde{g}_0(r)$, and the oscillations persist till large distances. The following sum rules hold (cf. Eq.~(\ref{eq-8}) for $n=1$):
\be
4\pi\int{\rm d}r\,r^2\rho\widetilde{g}(r)=\frac{1}{\rho}4\pi\int\frac{{\rm d}^3r_1}{V}\rho^{(1)}({\bf r}_1)\underbrace{\int\frac{{\rm d}^3r_2}{4\pi}\frac{\rho^{(2)}({\bf r}_1,{\bf r}_2)}{\rho^{(1)}({\bf r}_1)}}_{\frac{N-1}{4\pi}}=N-1
\label{eq-32}
\ee
and
\be
4\pi\int{\rm d}r\,r^2\rho\widetilde{g}_0(r)=\frac{1}{\rho}4\pi\int\frac{{\rm d}^3r_1}{V}\rho^{(1)}({\bf r}_1)\underbrace{\int\frac{{\rm d}^3r_2}{4\pi}\rho^{(1)}({\bf r}_2)}_{\frac{N}{4\pi}}=N\,.
\label{eq-33}
\ee
When the latter two formulae are rewritten as
\be
4\pi\rho\int{\rm d}r\,r^2(\widetilde{g}(r)-1)=-1\,\,\,\,\,\,{\rm and}\,\,\,\,\,\,4\pi\rho\int{\rm d}r\,r^2(\widetilde{g}_0(r)-1)=0\,,
\label{eq-34}
\ee
it becomes apparent that both $\widetilde{g}(r)$ and $\widetilde{g}_0(r)$ decay to 1 at infinity. Similarly, we define:
\be
\rho^2\widetilde{h}(r)=\int\frac{{\rm d}^3r_1}{V}\int\frac{{\rm d}^2\Omega}{4\pi}\,\rho^{(2)}({\bf r}_1,{\bf r}_1+{\bf r})\ln\frac{\rho^{(2)}({\bf r}_1,{\bf r}_1+{\bf r})}{\rho^{(1)}({\bf r}_1)\rho^{(1)}({\bf r}_1+{\bf r})}\,,
\label{eq-35}
\ee
which obviously vanishes at infinity. While $\widetilde{h}(r)=g(r)\ln g(r)$ for a homogeneous fluid, we expect that $\widetilde{h}(r)\neq\widetilde{g}(r)\ln\widetilde{g}(r)$ in the crystal. Putting Eqs.~(\ref{eq-30})-(\ref{eq-35}) together, we arrive at
\be
{\rm crystal}:\qquad S_2=-2\pi\rho N\int_0^\infty{\rm d}r\,r^2\left(\widetilde{h}(r)-\widetilde{g}(r)+\widetilde{g}_0(r)\right)=-2\pi\rho N\int_0^\infty{\rm d}r\,r^2\widetilde{h}(r)-\frac{N}{2}\,.
\label{eq-36}
\ee
Even though the integrand vanishes at infinity, $S_2={\cal O}(N)$ only if the {\em envelope} of $\widetilde{h}(r)$ decays faster than $r^{-3}$ ($r^{-2}$ in two dimensions). A slower decay may be sufficient if $S_2$ is computed through the first integral in (\ref{eq-36}). For a spherically-symmetric interaction potential, also the excess energy (i.e., the canonical average of the total potential energy $U$) can be written as a radial integral:
\ba
\left\langle U\right\rangle&=&\frac{1}{2}\int{\rm d}^3r_1{\rm d}^3r_2\,\rho^{(2)}({\bf r}_1,{\bf r}_2)u(|{\bf r}_2-{\bf r}_1|)
\nonumber \\
&=&\frac{1}{2}\int_0^\infty{\rm d}r\,r^2u(r)\int{\rm d}^3r_1\int{\rm d}^2\Omega\,\rho^{(2)}({\bf r}_1,{\bf r}_1+{\bf r})=2\pi\rho N\int_0^\infty{\rm d}r\,r^2u(r)\widetilde{g}(r)\,.
\label{eq-37}
\ea

For the one-body density in (\ref{eq-21}), $\widetilde{g}_0(r)$ can be obtained in closed form. First we have:
\be
\int\frac{{\rm d}^2\Omega}{4\pi}\,\rho^{(1)}({\bf r}_1+{\bf r})=\rho\sum_{\bf G}e^{-\frac{G^2}{4\alpha}}e^{i{\bf G}\cdot{\bf r}_1}\frac{\sin(Gr)}{Gr}\,.
\label{eq-38}
\ee
Then, multiplying by $\rho^{(1)}({\bf r}_1)=\rho\sum_{{\bf G}'}e^{-\frac{G^{\prime 2}}{4\alpha}}e^{i{\bf G}'\cdot{\bf r}_1}$ and finally integrating over ${\bf r}_1$ we arrive at
\be
\widetilde{g}_0(r)=1+\sum_{{\bf G}\neq 0}e^{-\frac{G^2}{2\alpha}}\frac{\sin(Gr)}{Gr}\,.
\label{eq-39}
\ee
We see that the large-distance decay of $\widetilde{g}_0(r)$ is usually slow, and the same will occur for $\widetilde{g}(r)$ since $\widetilde{g}(r)\simeq\widetilde{g}_0(r)$ for large $r$. In two dimensions, the one-body density and $\widetilde{g}_0$ functions respectively read:
\be
\rho^{(1)}({\bf r})=\frac{\alpha}{\pi}\sum_{\bf R}e^{-\alpha({\bf r}-{\bf R})^2}=\rho\sum_{\bf G}e^{-\frac{G^2}{4\alpha}}e^{i{\bf G}\cdot{\bf r}}\,\,\,{\rm and}\,\,\,\widetilde{g}_0(r)=1+\sum_{{\bf G}\neq 0}e^{-\frac{G^2}{2\alpha}}J_0(Gr)\,,
\label{eq-40}
\ee
where $J_0$ is a Bessel function of the first kind. Since the envelope of $J_0$ maxima decays as $r^{-1/2}$ at infinity, we see that the asymptotic vanishing of $\widetilde{g}_0$ is slower in two dimensions than in three.

Equation (\ref{eq-39}) has a well definite limit for $\alpha\rightarrow\infty$, corresponding to zero temperature. Indeed, using Poisson summation formula and the expression of Dirac's delta in spherical coordinates, we obtain:
\ba
\rho\widetilde{g}_0(r)&=&\rho\left(1+\sum_{{\bf G}\neq 0}\frac{\sin(Gr)}{Gr}\right)=\rho\int\frac{{\rm d}^2\Omega}{4\pi}\sum_{\bf G}e^{i{\bf G}\cdot{\bf r}}=\delta^3({\bf r})+\int\frac{{\rm d}^2\Omega}{4\pi}\sum_{{\bf R}\ne 0}\delta^3({\bf r}-{\bf R})
\nonumber \\
&=&\delta^3({\bf r})+\sum_{{\bf R}\ne 0}\frac{1}{4\pi}\int_0^{2\pi}{\rm d}\phi\int_0^\pi{\rm d}\theta\,\sin\theta\frac{1}{r^2\sin\theta}\delta(r-R)\delta(\theta-\theta_{\bf R})\delta(\phi-\phi_{\bf R})
\nonumber \\
&=&\delta^3({\bf r})+\sum_{{\bf R}\ne 0}\frac{1}{4\pi R^2}\delta(r-R)\,.
\label{eq-41}
\ea
Hence, $\widetilde{g}_0(r)$ reduces to a sum of delta functions centered at lattice distances (including the origin). The latter result is actually general. Inserting Eq.~(\ref{eq-23}) in (\ref{eq-31}), we obtain:
\small
\ba
\rho\widetilde{g}_0(r)&=&\frac{1}{\rho}\int\frac{{\rm d}^3r_1}{V}\int\frac{{\rm d}^2\Omega}{4\pi}\sum_{\bf R}\delta^3({\bf r}_1-{\bf R})\sum_{{\bf R}'}\delta^3({\bf r}_1+{\bf r}-{\bf R}')
\nonumber \\
&=&\frac{1}{\rho}\int\frac{{\rm d}^3r_1}{V}\int\frac{{\rm d}^2\Omega}{4\pi}\left\{\sum_{\bf R}\delta^3({\bf r}_1-{\bf R})\delta^3({\bf r}_1+{\bf r}-{\bf R})+\sum_{{\bf R}\ne{\bf R}'}\delta^3({\bf r}_1-{\bf R})\delta^3({\bf r}_1+{\bf r}-{\bf R}')\right\}
\nonumber \\
&=&\frac{1}{\rho}\int\frac{{\rm d}^3r_1}{V}\int\frac{{\rm d}^2\Omega}{4\pi}\left\{\sum_{\bf R}\delta^3({\bf r}_1-{\bf R})\delta^3({\bf r})+\sum_{{\bf R}\ne{\bf R}'}\delta^3({\bf r}_1-{\bf R})\delta^3({\bf r}_1+{\bf r}-{\bf R}')\right\}
\nonumber \\
&=&\frac{1}{\rho}\sum_{\bf R}\delta^3({\bf r})\int\frac{{\rm d}^3r_1}{V}\delta^3({\bf r}_1-{\bf R})+\frac{1}{\rho}\sum_{{\bf R}\ne{\bf R}'}\int\frac{{\rm d}^3r_1}{V}\delta^3({\bf r}_1-{\bf R})\frac{\delta(r-|{\bf r}_1-{\bf R}'|)}{4\pi|{\bf r}_1-{\bf R}'|^2}
\nonumber \\
&=&\delta^3({\bf r})+\sum_{{\bf R}\ne 0}\frac{1}{4\pi R^2}\delta(r-R)\,,
\label{eq-42}
\ea
\normalsize
q.e.d. At zero temperature, $\rho\widetilde{g}(r)$ is given by the same sum of delta-function terms as in (\ref{eq-42}), {\em but} for the first term, $\delta^3({\bf r})$, which is missing --- see Eq.\,(\ref{eq-92}) below.

We add a final comment on possible alternative formulations of $\widetilde{g}(r)$ for a crystal. One choice is to replace (\ref{eq-30}) with
\be
{\rm option\,\,B}:\qquad\rho\widetilde{g}(r)=\int\frac{{\rm d}^3r_1}{V}\int\frac{{\rm d}^2\Omega}{4\pi}\frac{\rho^{(2)}({\bf r}_1,{\bf r}_1+{\bf r})}{\rho^{(1)}({\bf r}_1)}\,.
\label{eq-43}
\ee
Apparently, this is a good definition since (see Eq.~(\ref{eq-8}))
\be
4\pi\int{\rm d}r\,r^2\rho\widetilde{g}(r)=4\pi\underbrace{\int\frac{{\rm d}^3r_1}{V}}_{1}\underbrace{\int\frac{{\rm d}^3r_2}{4\pi}\frac{\rho^{(2)}({\bf r}_1,{\bf r}_2)}{\rho^{(1)}({\bf r}_1)}}_{\frac{N-1}{4\pi}}=N-1\,.
\label{eq-44}
\ee
However, with this $\widetilde{g}(r)$ we cannot write $S_2$ as a radial integral --- hence option B is discarded altogether. Another possibility is
\be
{\rm option\,\,C}:\qquad\widetilde{g}(r)=\int\frac{{\rm d}^3r_1}{V}\int\frac{{\rm d}^2\Omega}{4\pi}\frac{\rho^{(2)}({\bf r}_1,{\bf r}_1+{\bf r})}{\rho^{(1)}({\bf r}_1)\rho^{(1)}({\bf r}_1+{\bf r})}\,,
\label{eq-45}
\ee
but this option is useless too, since
\be
4\pi\int{\rm d}r\,r^2\rho\widetilde{g}(r)=\rho\int{\rm d}^3r_1\frac{1}{V}\int{\rm d}^3r_2\,g^{(2)}({\bf r}_1,{\bf r}_2)=\,?
\label{eq-46}
\ee
(observe that the inner integral is different from the one appearing in Eq.~(\ref{eq-8})).

\subsection{Symmetries of the two-body density}
A general property of the two-body density for a crystal is the CE sum rule
\be
\int{\rm d}^3r_2\,\rho^{(2)}({\bf r}_1,{\bf r}_2)=(N-1)\rho^{(1)}({\bf r}_1)\,.
\label{eq-47}
\ee
Other constraints follow from the translational symmetry of local crystal properties. Likewise the one-body density, fulfilling $\rho^{(1)}({\bf r}_1+{\bf R})=\rho^{(1)}({\bf r}_1)$ for every ${\bf R}$, we must have that
\be
\rho^{(2)}({\bf r}_1,{\bf r}_2)=\rho^{(2)}({\bf r}_1+{\bf R},{\bf r}_2+{\bf R})\,,
\label{eq-48}
\ee
in turn implying
\be
g^{(2)}({\bf r}_1,{\bf r}_2)=g^{(2)}({\bf r}_1+{\bf R},{\bf r}_2+{\bf R})\,.
\label{eq-49}
\ee
Now observe \cite{Gernoth} that i) any function of ${\bf r}_1$ and ${\bf r}_2$ can also be viewed as a function of $({\bf r}_1+{\bf r}_2)/2$ and ${\bf r}_2-{\bf r}_1$; ii) under a ${\bf R}$-translation, only the former variable is affected, not the relative separation. Hence, the most general function consistent with (\ref{eq-49}) is:
\be
g^{(2)}({\bf r}_1,{\bf r}_2)=\sum_{\bf G}\widetilde{v}_{\bf G}({\bf r}_2-{\bf r}_1)e^{i{\bf G}\cdot\frac{{\bf r}_1+{\bf r}_2}{2}}\,,
\label{eq-50}
\ee
where
\be
g^{(2)}({\bf r}_1,{\bf r}_2)\in\mathbb{R}\Longrightarrow\widetilde{v}_{\bf G}^*({\bf r}_2-{\bf r}_1)=\widetilde{v}_{-{\bf G}}({\bf r}_2-{\bf r}_1)
\label{eq-51}
\ee
and
\be
g^{(2)}({\bf r}_1,{\bf r}_2)=g^{(2)}({\bf r}_2,{\bf r}_1)\Longrightarrow\widetilde{v}_{\bf G}({\bf r}_2-{\bf r}_1)=\widetilde{v}_{\bf G}({\bf r}_1-{\bf r}_2)\,.
\label{eq-52}
\ee
In order that $\lim_{r\rightarrow\infty}g^{(2)}({\bf r}_1,{\bf r}_1+{\bf r})=1$ it is sufficient that
\be
\lim_{r\rightarrow\infty}\widetilde{v}_0({\bf r})=1\,\,\,\,\,\,{\rm and}\,\,\,\,\,\,\lim_{r\rightarrow\infty}\widetilde{v}_{\bf G}({\bf r})=0\,\,{\rm for}\,\,{\bf G}\neq 0\,.
\label{eq-53}
\ee
We may reasonably expect that the most relevant term in the expansion (\ref{eq-50}) is indeed the ${\bf G}=0$ one (also notice that $\widetilde{v}_{\bf G}\rightarrow 0$ as $G\rightarrow\infty$ by the Riemann-Lebesgue lemma).

Equation (\ref{eq-50}) is still insufficient to establish the scaling of two-body entropy with the size of the crystal. Some general results can be obtained under the (strong) {\em assumption} that $\widetilde{v}_{\bf G}({\bf r})=0$ for any ${\bf G}\neq 0$. If we change the notation from $\widetilde{v}_0$ to ${\cal G}({\bf r})\equiv 1+{\cal H}({\bf r})$ (which, by Eqs.~(\ref{eq-51}) and (\ref{eq-52}), is a real and even function), then a necessary condition for ${\cal H}$ is:
\be
\int{\rm d}^3r_2\,\rho^{(1)}({\bf r}_2){\cal H}({\bf r}_2-{\bf r}_1)=-1\,\,\,\,\,\,{\rm for\,\,any}\,\,{\bf r}_1\,\,{\rm where}\,\,\rho^{(1)}({\bf r}_1)\neq 0\,.
\label{eq-54}
\ee
The rationale behind Eq.~(\ref{eq-54}) is particularly transparent near $T=0$, where the peaks of the one-body density are extremely narrow. As argued below (see Eq.~(\ref{eq-67}) ff.), ${\cal H}$ as a function of ${\bf r}_2$ is roughly $-1$ in the primitive cell ${\cal C}$ centered in ${\bf r}_1\approx{\bf R}_1$, denoting ${\bf R}_1$ the only lattice site contained in ${\cal C}$, and roughly zero outside ${\cal C}$. Since the integral of $\rho^{(1)}$ over ${\cal C}$ equals 1, Eq.~(\ref{eq-54}) will immediately follow.

Now writing ${\cal H}({\bf r})$ as a Fourier integral,
\be
{\cal H}({\bf r})=\int\frac{{\rm d}^3k}{(2\pi)^3}\widetilde{\cal H}({\bf k})e^{i{\bf k}\cdot{\bf r}}\,,
\label{eq-55}
\ee
and using (\ref{eq-21}) as one-body density, Eq.~(\ref{eq-54}) yields
\be
\rho\sum_{\bf G}e^{-\frac{G^2}{4\alpha}}\widetilde{\cal H}({\bf G})e^{-i{\bf G}\cdot{\bf r}_1}=-1\,,
\label{eq-56}
\ee
which can only hold for arbitrary ${\bf r}_1$ if
\be
\widetilde{\cal H}({\bf G})=-\frac{1}{\rho}\delta_{{\bf G},0}\,.
\label{eq-57}
\ee
Next, from Eq.~(\ref{eq-30}) we obtain:
\be
\rho^2\widetilde{g}(r)=\rho^2\widetilde{g}_0(r)+\int\frac{{\rm d}^3r_1}{V}\rho^{(1)}({\bf r}_1)\int\frac{{\rm d}^2\Omega}{4\pi}\,\rho^{(1)}({\bf r}_1+{\bf r}){\cal H}({\bf r})\,.
\label{eq-58}
\ee
For the one-body density in (\ref{eq-21}), the inner integral becomes:
\be
\int\frac{{\rm d}^2\Omega}{4\pi}\,\rho^{(1)}({\bf r}_1+{\bf r}){\cal H}({\bf r})=\rho\sum_{\bf G}e^{-\frac{G^2}{4\alpha}}I_{\bf G}(r)e^{-i{\bf G}\cdot{\bf r}_1}
\label{eq-59}
\ee
with
\be
I_{\bf G}(r)=\int\frac{{\rm d}^2\Omega}{4\pi}\,{\cal H}({\bf r})e^{-i{\bf G}\cdot{\bf r}}=\int\frac{{\rm d}^3k}{(2\pi)^3}\,\widetilde{\cal H}({\bf k})\frac{\sin\left(|{\bf k}-{\bf G}|r\right)}{|{\bf k}-{\bf G}|r}\,.
\label{eq-60}
\ee
It is evident that $I_{\bf G}(r)$ vanishes at infinity. Upon inserting (\ref{eq-59}) in (\ref{eq-58}), we finally obtain:
\be
\widetilde{g}(r)=\widetilde{g}_0(r)+\sum_{\bf G}e^{-\frac{G^2}{2\alpha}}I_{\bf G}(r)\,.
\label{eq-61}
\ee
As $r$ increases, the second term gradually vanishes and the large-distance oscillations of $\widetilde{g}(r)$ then exactly match those of $\widetilde{g}_0(r)$. As a countercheck, let us compute the integral of $\rho\widetilde{g}(r)-\rho\widetilde{g}_0(r)$ over the macroscopic system volume (which, by Eqs.~(\ref{eq-32}) and (\ref{eq-33}), should be $-1$):
\ba
4\pi\int{\rm d}r\,r^2\rho(\widetilde{g}(r)-\widetilde{g}_0(r))&=&\rho\sum_{\bf G}e^{-\frac{G^2}{2\alpha}}\cdot 4\pi\int{\rm d}r\,r^2I_{\bf G}(r)
\nonumber \\
&=&\rho\sum_{\bf G}e^{-\frac{G^2}{2\alpha}}\int\frac{{\rm d}^3k}{(2\pi)^3}\widetilde{\cal H}({\bf k})\cdot 4\pi\int{\rm d}r\,r^2\frac{\sin\left(|{\bf k}-{\bf G}|r\right)}{|{\bf k}-{\bf G}|r}
\nonumber \\
&=&\rho\sum_{\bf G}e^{-\frac{G^2}{2\alpha}}\int\frac{{\rm d}^3k}{(2\pi)^3}\widetilde{\cal H}({\bf k})\underbrace{\int{\rm d}^3r\,e^{i({\bf k}-{\bf G})\cdot{\bf r}}}_{(2\pi)^3\delta^3({\bf k}-{\bf G})}
\nonumber \\
&=&\rho\sum_{\bf G}e^{-\frac{G^2}{2\alpha}}\underbrace{\widetilde{\cal H}({\bf G})}_{-(1/\rho)\delta_{{\bf G},0}}=-1\,.
\label{eq-62}
\ea

Under the assumption that
\be
\rho^{(2)}({\bf r}_1,{\bf r}_1+{\bf r})=\rho^{(1)}({\bf r}_1)\rho^{(1)}({\bf r}_1+{\bf r}){\cal G}({\bf r})\,,
\label{eq-63}
\ee
the entropy expansion for a crystal reads
\ba
\frac{S}{k_B}=&&-\int{\rm d}^3r_1\,\rho^{(1)}({\bf r}_1)\ln\frac{\rho^{(1)}({\bf r}_1)}{\rho}
\nonumber \\
&&-\frac{1}{2}\int{\rm d}^3r_1\,\rho^{(1)}({\bf r}_1)\int{\rm d}^3r\,\rho^{(1)}({\bf r}_1+{\bf r})\left[{\cal G}({\bf r})\ln{\cal G}({\bf r})-{\cal G}({\bf r})+1\right]+\ldots
\label{eq-64}
\ea
Providing that it vanishes sufficiently rapidly at infinity, the function
\be
{\cal K}({\bf r})={\cal G}({\bf r})\ln{\cal G}({\bf r})-{\cal G}({\bf r})+1
\label{eq-65}
\ee
can be written as a Fourier integral and, using (\ref{eq-21}) as one-body density, the two-body entropy becomes
\be
S_2=-\frac{1}{2}\rho^2\sum_{{\bf G},{\bf G}'}e^{-\frac{G^2+G^{\prime 2}}{4\alpha}}\underbrace{\int{\rm d}^3r_1\,e^{i({\bf G}+{\bf G}')\cdot{\bf r}_1}}_{V\delta_{{\bf G}',-{\bf G}}}\int{\rm d}^3r\,{\cal K}({\bf r})e^{i{\bf G}'\cdot{\bf r}}=-\frac{1}{2}N\rho\sum_{\bf G}e^{-\frac{G^2}{2\alpha}}\widetilde{\cal K}({\bf G})\,,
\label{eq-66}
\ee
which is clearly ${\cal O}(N)$.

\subsection{Two-body density at $T=0$}

In the zero-temperature limit, particles will be sitting at lattice sites, and the two-body density then becomes (see Eq.~(\ref{eq-23})):
\be
\rho^{(2)}({\bf r}_1,{\bf r}_2)=\sideset{}{^\prime}\sum_{{\bf R},{\bf R}'}\delta^3({\bf r}_1-{\bf R})\delta^3({\bf r}_2-{\bf R}')=\rho^{(1)}({\bf r}_1)\rho^{(1)}({\bf r}_2)\left(1-{\bf 1}_{{\cal C}}({\bf r}_2-{\bf r}_1)\right)\,,
\label{eq-67}
\ee
which is of the form (\ref{eq-63}). In Eq.~(\ref{eq-67}), ${\bf 1}_{{\cal C}}({\bf r})$ is the indicator function of a Wigner-Seitz cell ${\cal C}$ centered at the origin (i.e., ${\bf 1}_{{\cal C}}({\bf r})=1$ if ${\bf r}\in{\cal C}$ and ${\bf 1}_{{\cal C}}({\bf r})=0$ otherwise). While the factor $\rho^{(1)}({\bf r}_1)\rho^{(1)}({\bf r}_2)$ forces particles to be located at lattice sites, the only role of the ${\cal G}$ in (\ref{eq-67}) is to prevent the possibility of double site occupancy. However, a ${\cal G}$ function with this property is not unique; the one provided in (\ref{eq-67}) has the advantage of exactly complying with condition (\ref{eq-57}) (see below). Equation (\ref{eq-67}) indicates that the pair-correlation structure of a low-temperature solid is very different from the structure of a dense fluid close to freezing.

For
\be
{\cal H}({\bf r})=-{\bf 1}_{{\cal C}}({\bf r})=\left\{
\begin{array}{rl}
-1, & \,\,\,{\rm for}\,\,{\bf r}\in{\cal C}\\
0, & \,\,\,{\rm otherwise}
\end{array}
\right.
\label{eq-68}
\ee
the Fourier transform reads:
\ba
\widetilde{\cal H}({\bf k})&=&\int{\rm d}^3r\,{\cal H}({\bf r})e^{-i{\bf k}\cdot{\bf r}}=-\int_{\cal C}{\rm d}^3r\,e^{-i{\bf k}\cdot{\bf r}}\,.
\label{eq-69}
\ea
Now observe that $f({\bf r})=1$ is trivially periodic, and can thus be expanded in plane waves as $1=\sum_{\bf G}\widetilde{f}_{\bf G}e^{i{\bf G}\cdot{\bf r}}$, with $\widetilde{f}_{\bf G}=\delta_{{\bf G},0}$. On the other hand,
\be
\widetilde{f}_{\bf G}=\frac{1}{v_0}\int_{\cal C}{\rm d}^3r\,f({\bf r})e^{-i{\bf G}\cdot{\bf r}}=\rho\int_{\cal C}{\rm d}^3r\,e^{-i{\bf G}\cdot{\bf r}}\,.
\label{eq-70}
\ee
Comparing Eqs.~(\ref{eq-69}) and (\ref{eq-70}), we conclude that
\be
\widetilde{\cal H}({\bf G})=-\frac{1}{\rho}\delta_{{\bf G},0}\,.
\label{eq-71}
\ee

For ${\cal H}({\bf r})=-{\bf 1}_{{\cal C}}({\bf r})$ the function $I_{\bf G}(r)$ at Eq.~(\ref{eq-60}) equals $-\sin(Gr)/(Gr)$ for $r<r_m$ and 0 for $r>r_M$, where $r_m$ ($r_M$) is the radius of the largest (smallest) sphere inscribed in (circumscribed to) ${\cal C}$. It then follows from Eq.~(\ref{eq-61}) that $\widetilde{g}(r)=0$ for $r<r_m$, while $\widetilde{g}(r)=\widetilde{g}_0(r)$ for $r>r_M$ (for a triangular crystal with spacing $a$ we have $r_m=a/2$ and $r_M=a/\sqrt{3}$, both comprised between the first, 0, and the second, $a$, lattice distance). For $T=0$, where $\widetilde{g}_0(r)$ consists of infinitely narrow peaks centered at lattice distances, this implies that $\widetilde{g}(r)=\widetilde{g}_0(r)$ everywhere but at the origin, where $\widetilde{g}(r)=0$ while $\widetilde{g}_0(r)$ is non-zero.

\subsection{Scaling of two-body entropy with $N$}

We henceforth discuss in fully general terms how the two-body entropy scales with $N$ for a crystal, avoiding to make any simplifying hypothesis on the structure of $g^{(2)}({\bf r}_1,{\bf r}_2)$. Using an obvious short-hand notation, the two-body entropy reads
\be
S_2=-\frac{1}{2}\int{\rm d}1\,{\rm d}2\left(\rho_{12}\ln\frac{\rho_{12}}{\rho_1\rho_2}-\rho_{12}+\rho_1\rho_2\right)=-\frac{1}{2}\int{\rm d}1\,{\rm d}2\,\rho_1\rho_2\left(g_{12}\ln g_{12}-g_{12}+1\right)\,.
\label{eq-72}
\ee
As we already know, $S_2\le 0$. From the inequality $\ln x\le x-1$, valid for all $x>0$, we derive $-x\ln x\ge x-x^2$ for $x\ge 0$, and then obtain:
\be
S_2=\frac{1}{2}\int{\rm d}1\,{\rm d}2\,\rho_1\rho_2\left(-g_{12}\ln g_{12}+g_{12}-1\right)\ge-\frac{1}{2}\int{\rm d}1\,{\rm d}2\,\rho_1\rho_2\left(g_{12}-1\right)^2\,.
\label{eq-73}
\ee
Clearly, estimating the size of the lower bound in Eq.~(\ref{eq-73}) is a much simpler problem than working with $S_2$ itself.

Taking $h_{12}\equiv g_{12}-1$, it is evident that $h_{12}$ shares all symmetries of $g_{12}$. Hence, we can write:
\be
h_{12}=\sum_{\bf G}\widetilde{h}_{\bf G}({\bf r}_1-{\bf r}_2)e^{i{\bf G}\cdot\frac{{\bf r}_1+{\bf r}_2}{2}}\,\,\,\,\,\,{\rm with}\,\,\,\,\,\,\widetilde{h}_{\bf G}^*({\bf r})=\widetilde{h}_{-{\bf G}}({\bf r})\,\,\,{\rm and}\,\,\,\widetilde{h}_{\bf G}({\bf r})=\widetilde{h}_{\bf G}(-{\bf r})\,.
\label{eq-74}
\ee
Observe that the $\widetilde{h}_{\bf G}({\bf r})$ functions are nothing but Fourier coefficients, once the $h$ function has been expressed in terms of ${\bf S}=({\bf r}_1+{\bf r}_2)/2$ and ${\bf r}={\bf r}_1-{\bf r}_2$:
\be
\widetilde{h}_{\bf G}({\bf r})=\frac{1}{v_0}\int_{\cal C}{\rm d}^3S\,h({\bf S}+{\bf r}/2,{\bf S}-{\bf r}/2)e^{-i{\bf G}\cdot{\bf S}}\,.
\label{eq-75}
\ee
By the Riemann-Lebesgue lemma, $\widetilde{h}_{\bf G}({\bf r})\rightarrow 0$ as $G\rightarrow\infty$ (for arbitrary ${\bf r}$). Moreover, $\widetilde{h}_{\bf G}({\bf r})\rightarrow 0$ for $r\rightarrow\infty$ (for arbitrary ${\bf G}$) since $\rho_{12}\rightarrow\rho_1\rho_2$ for $|{\bf r}_1-{\bf r}_2|\rightarrow\infty$. Similarly, for $k_{12}\equiv h_{12}^2$ we have that
\be
k_{12}=\sum_{\bf G}\widetilde{k}_{\bf G}({\bf r}_1-{\bf r}_2)e^{i{\bf G}\cdot\frac{{\bf r}_1+{\bf r}_2}{2}}\,\,\,\,\,\,{\rm with}\,\,\,\,\,\,\widetilde{k}_{\bf G}({\bf r})=\sum_{{\bf G}'}\widetilde{h}_{{\bf G}-{\bf G}'}({\bf r})\widetilde{h}_{{\bf G}'}({\bf r})\,.
\label{eq-76}
\ee
Now observe that, for $\rho^{(1)}({\bf r})=\sum_{\bf G}\widetilde{u}_{\bf G}e^{i{\bf G}\cdot{\bf r}}$,
\be
\rho^{(1)}({\bf r})\rho^{(1)}({\bf r}')=\sum_{\bf G}\underbrace{\left(\sum_{{\bf G}'}\widetilde{u}_{{\bf G}-{\bf G}'}\widetilde{u}_{{\bf G}'}e^{i(2{\bf G}'-{\bf G})\cdot\frac{{\bf r}-{\bf r}'}{2}}\right)}_{\widetilde{v}_{\bf G}^\infty({\bf r}-{\bf r}')}e^{i{\bf G}\cdot\frac{{\bf r}+{\bf r}'}{2}}\,.
\label{eq-77}
\ee
Using the above equation, and changing the integration variables from ${\bf r}_1$ and ${\bf r}_2$ to ${\bf S}$ and ${\bf r}$, we obtain:
\ba
-\frac{1}{2}\int{\rm d}1\,{\rm d}2\,\rho_1\rho_2\left(g_{12}-1\right)^2&=&-\frac{1}{2}\int{\rm d}^3S\,{\rm d}^3r\sum_{\bf G}\widetilde{v}_{\bf G}^\infty({\bf r})e^{i{\bf G}\cdot{\bf S}}\sum_{{\bf G}'}\widetilde{k}_{{\bf G}'}({\bf r})e^{i{\bf G}'\cdot{\bf S}}
\nonumber \\
&=&-\frac{1}{2}\sum_{{\bf G},{\bf G}'}\underbrace{\int{\rm d}^3S\,e^{i({\bf G}+{\bf G}')\cdot{\bf S}}}_{V\delta_{{\bf G}',-{\bf G}}}\underbrace{\int{\rm d}^3r\,\widetilde{v}_{\bf G}^\infty({\bf r})\widetilde{k}_{{\bf G}'}({\bf r})}_{i_{{\bf G},{\bf G}'}}
\nonumber \\
&=&-\frac{1}{2}V\sum_{\bf G}i_{{\bf G},-{\bf G}}\,,
\label{eq-78}
\ea
where
\be
i_{{\bf G},-{\bf G}}=\sum_{{\bf G}'}\widetilde{u}_{{\bf G}-{\bf G}'}\widetilde{u}_{{\bf G}'}\int{\rm d}^3r\left(\sum_{{\bf G}''}\widetilde{h}_{-{\bf G}-{\bf G}''}({\bf r})\widetilde{h}_{{\bf G}''}({\bf r})\right)e^{i(2{\bf G}'-{\bf G})\cdot\frac{\bf r}{2}}\,.
\label{eq-79}
\ee

In the special case $\widetilde{h}_{\bf G}={\cal H}({\bf r})\delta_{{\bf G},0}$, we have $h_{12}=\widetilde{h}_0={\cal H}({\bf r}_1-{\bf r}_2)$ and $\widetilde{k}_{\bf G}({\bf r})={\cal H}({\bf r})^2\delta_{{\bf G},0}$. Then, from Eq.~(\ref{eq-77}) we derive
\be
\sum_{\bf G}i_{{\bf G},-{\bf G}}=i_{0,0}=\int{\rm d}^3r\,\widetilde{v}_0^\infty({\bf r})\widetilde{k}_0({\bf r})=\sum_{\bf G}\left|\widetilde{u}_{\bf G}\right|^2\widetilde{{\cal H}^2}({\bf G})\,.
\label{eq-80}
\ee
An independent computation of the integral leads to the same result:
\ba
-\frac{1}{2}\int{\rm d}1\,{\rm d}2\,\rho_1\rho_2\left(g_{12}-1\right)^2&=&-\frac{1}{2}\int{\rm d}^3r\,\rho^{(1)}({\bf r})\int{\rm d}^3r'\,\underbrace{\rho^{(1)}({\bf r}+{\bf r}')}_{\sum_{\bf G}\widetilde{u}_{\bf G}^*e^{-i{\bf G}\cdot{({\bf r}+{\bf r}')}}}{\cal H}^2({\bf r}')
\nonumber \\
&=&-\frac{1}{2}\sum_{\bf G}\widetilde{u}_{\bf G}^*\underbrace{\int{\rm d}^3r\,\rho^{(1)}({\bf r})e^{-i{\bf G}\cdot{\bf r}}}_{V\widetilde{u}_{\bf G}}\underbrace{\int{\rm d}^3r'\,{\cal H}^2({\bf r}')e^{-i{\bf G}\cdot{\bf r}'}}_{\widetilde{{\cal H}^2}({\bf G})}
\nonumber \\
&=&-\frac{1}{2}V\sum_{\bf G}\left|\widetilde{u}_{\bf G}\right|^2\widetilde{{\cal H}^2}({\bf G})\,,
\label{eq-81}
\ea
which should be compared with Eq.~(\ref{eq-66}). For ${\cal H}({\bf r})=-{\bf 1}_{{\cal C}}({\bf r})$ and $\widetilde{u}_{\bf G}=\rho\exp\{-G^2/(4\alpha)\}$, we readily obtain $S_2=-N/2$ from both Eqs.~(\ref{eq-66}) and (\ref{eq-78}), meaning that in this case the two-body entropy coincides with its lower bound in Eq.~(\ref{eq-73}).

The quantity (\ref{eq-80}) is clearly ${\cal O}(1)$, since the summand is rapidly converging to zero; this implies that the two-body entropy of a crystal is, at least for $\widetilde{h}_{\bf G}={\cal H}({\bf r})\delta_{{\bf G},0}$, bounded from below by a ${\cal O}(N)$ quantity. In the most general case, where Eqs.~(\ref{eq-78}) and (\ref{eq-79}) rather apply, we can only observe the following. As $G$ grows in size, for any fixed ${\bf G}'$ and ${\bf G}''$ both $\widetilde{u}_{{\bf G}-{\bf G}'}\widetilde{u}_{{\bf G}'}$ and $\widetilde{h}_{-{\bf G}-{\bf G}''}({\bf r})\widetilde{h}_{{\bf G}''}({\bf r})$ get smaller, suggesting that $i_{{\bf G},-{\bf G}}$ will decrease too. However, this is not enough to conclude that $\sum_{\bf G}i_{{\bf G},-{\bf G}}$ is ${\cal O}(1)$, and the only way to settle the problem is numerical.

\subsection{Numerical evaluation of the structure functions}
The utility of (\ref{eq-36}) clearly relies on the possibility of determining the integrand in simulation with sufficient accuracy. First we see how the one-body entropy, Eq.~(\ref{eq-24}), is computed. We start dividing $V$ into a large number $M=V/v_c$ of identical cubes of volume $v_c$, chosen to be small enough that a cube contains the center of at most one particle. Let $c_\alpha=0,1$ (with $\alpha=1,\ldots,M$) be the occupancy of the $\alpha$th cube in a given system configuration and $\langle c_\alpha\rangle$ its canonical average as computed in a long Monte Carlo simulation of the {\em weakly constrained} crystal (to fix the center of mass of the crystal in space it is sufficient to keep one particle fixed; then, periodic boundary conditions will contribute to keep crystalline axes also fixed in the course of simulation). Given this setup, the local density at ${\bf r}_1$ (a point inside the $\alpha$th cube) can be estimated as
\be
\rho^{(1)}({\bf r}_1)\approx\frac{\left\langle c_\alpha\right\rangle}{v_c}\,,
\label{eq-82}
\ee
and the integral in (\ref{eq-24}) becomes
\be
\int{\rm d}^3r_1\,\rho^{(1)}({\bf r}_1)\ln\frac{\rho^{(1)}({\bf r}_1)}{\rho}\approx\sum_{\alpha=1}^M\left\langle c_\alpha\right\rangle\ln\frac{\left\langle c_\alpha\right\rangle}{\rho v_c}
\label{eq-83}
\ee
(notice that $\rho v_c=N/M\ll 1$; we need $v_c\rightarrow 0$ and an infinitely long simulation to make (\ref{eq-82}) an exact relation). Similarly, if ${\bf r}_2$ falls within the $\beta$th cube, then
\be
\rho^{(2)}({\bf r}_1,{\bf r}_2)\approx\frac{\left\langle c_\alpha c_\beta\right\rangle}{v_c^2}
\label{eq-84}
\ee
and from Eq.~(\ref{eq-30}) we derive
\ba
\widetilde{g}(r)&\approx&\frac{1}{\rho^2V}\sum_{\alpha=1}^Mv_c\frac{1}{N_\gamma}\sum_{|\gamma|=r}\frac{\left\langle c_\alpha c_{\alpha+\gamma}\right\rangle}{v_c^2}=\frac{1}{\rho^2V}\left\langle\sum_{\alpha=1}^M v_c\frac{1}{N_\gamma}\sum_{|\gamma|=r}\frac{c_\alpha c_{\alpha+\gamma}}{v_c^2}\right\rangle
\nonumber \\
&=&\frac{1}{M\rho^2 v_c^2}\left\langle\sum_{\alpha=1}^M\delta_{c_\alpha,1}\frac{1}{N_\gamma}\sum_{|\gamma|=r}c_{\alpha+\gamma}\right\rangle\,.
\label{eq-85}
\ea
In the above formula $N_\gamma\simeq 4\pi r^2\Delta r/v_c$ is the number of cubes whose center lies at a distance $r$ from $\alpha$ (to within a certain tolerance $\Delta r\ll r$) and the inner sum is carried out over those cubes only. Since $\rho v_cN_\gamma=4\pi r^2\Delta r\rho$ and $M\rho v_c=N$, an equivalent formula for $\widetilde{g}(r)$ is
\be
\widetilde{g}(r)\approx\left\langle\frac{1}{N}\sum_{i=1}^N\frac{{\cal N}_i(r\pm\Delta r/2)}{4\pi r^2\Delta r\rho}\right\rangle\,,
\label{eq-86}
\ee
denoting ${\cal N}_i(r\pm\Delta r/2)$ the number of particles found at a distance between $r-\Delta r/2$ and $r+\Delta r/2$ from the $i$th particle in the given configuration. Equation (\ref{eq-86}) closely reflects the method of computing the radial distribution function in a CE simulation (see, e.g., Eq.~(11) in Ref.~\cite{Prestipino7}).

The function $\widetilde{g}(r)$ admits yet another expression, which further strengthens its resemblance to the $g(r)$ of a liquid (as reported e.g. in \cite{Hansen}). It follows from Eqs.~(\ref{eq-30}) and (\ref{eq-6}) that
\be
\rho^2\widetilde{g}(r)=\frac{1}{V}\int\frac{{\rm d}^2\Omega}{4\pi}\left\langle\sideset{}{^\prime}\sum_{ij}\int{\rm d}^3r_1\,\delta^3({\bf r}_1-{\bf R}_i)\delta^3({\bf r}_1+{\bf r}-{\bf R}_j)\right\rangle\,.
\label{eq-87}
\ee
Observing that, for any sufficiently smooth function $f({\bf r})$,
\small
\ba
\int{\rm d}^3r\,f({\bf r})\int{\rm d}^3r_1\,\delta^3({\bf r}_1-{\bf R}_i)\delta^3({\bf r}_1+{\bf r}-{\bf R}_j)&=&\int{\rm d}^3r_1\,\delta^3({\bf r}_1-{\bf R}_i)\int{\rm d}^3r\,f({\bf r})\delta^3({\bf r}_1+{\bf r}-{\bf R}_j)
\nonumber \\
&=&\int{\rm d}^3r_1\,\delta^3({\bf r}_1-{\bf R}_i)f({\bf R}_j-{\bf r}_1)=f({\bf R}_j-{\bf R}_i)
\label{eq-88}
\ea
\normalsize
and
\small
\ba
\int{\rm d}^3r\,f({\bf r})\int{\rm d}^3r_1\,\delta^3({\bf r}_1-{\bf R}_i)\delta^3({\bf R}_i+{\bf r}-{\bf R}_j)&=&\int{\rm d}^3r_1\,\delta^3({\bf r}_1-{\bf R}_i)\int{\rm d}^3r\,f({\bf r})\delta^3({\bf R}_i+{\bf r}-{\bf R}_j)
\nonumber \\
&=&f({\bf R}_j-{\bf R}_i)\underbrace{\int{\rm d}^3r_1\,\delta^3({\bf r}_1-{\bf R}_i)}_{1}=f({\bf R}_j-{\bf R}_i)\,,
\label{eq-89}
\ea
\normalsize
we are allowed to replace $\delta^3({\bf r}_1-{\bf R}_i)\delta^3({\bf r}_1+{\bf r}-{\bf R}_j)$ with $\delta^3({\bf r}_1-{\bf R}_i)\delta^3({\bf R}_i+{\bf r}-{\bf R}_j)$ in Eq.\,(\ref{eq-87}), and thus obtain
\be
\rho^2\widetilde{g}(r)=\frac{1}{V}\int\frac{{\rm d}^2\Omega}{4\pi}\left\langle\sideset{}{^\prime}\sum_{ij}\delta^3({\bf R}_i+{\bf r}-{\bf R}_j)\underbrace{\int{\rm d}^3r_1\,\delta^3({\bf r}_1-{\bf R}_i)}_{1}\right\rangle=\frac{1}{V}\int\frac{{\rm d}^2\Omega}{4\pi}\left\langle\sideset{}{^\prime}\sum_{ij}\delta^3({\bf R}_i+{\bf r}-{\bf R}_j)\right\rangle\,,
\label{eq-90}
\ee
which finally leads to
\be
\rho\widetilde{g}(r)=\int\frac{{\rm d}^2\Omega}{4\pi}\left\langle\frac{1}{N}\sum_i\sum_{j\neq i}\delta^3({\bf R}_i+{\bf r}-{\bf R}_j)\right\rangle\,.
\label{eq-91}
\ee
At zero temperature, we can neglect the average and simply write
\be
\rho\widetilde{g}(r)=\int\frac{{\rm d}^2\Omega}{4\pi}\frac{1}{N}\sum_i\sum_{j\neq i}\delta^3({\bf R}_i+{\bf r}-{\bf R}_j)=\int\frac{{\rm d}^2\Omega}{4\pi}\sum_{{\bf R}\ne 0}\delta^3({\bf r}-{\bf R})=\sum_{{\bf R}\ne 0}\frac{1}{4\pi R^2}\delta(r-R)\,,
\label{eq-92}
\ee
where in the last step we have followed the same path leading to Eq.~(\ref{eq-41}).

We can similarly proceed for the functions at Eqs.~(\ref{eq-31}) and (\ref{eq-35}), which can be computed by the following formulae:
\be
\widetilde{g}_0(r)\approx\frac{1}{\rho^2V}\sum_{\alpha=1}^Mv_c\frac{1}{N_\gamma}\sum_{|\gamma|=r}\frac{\left\langle c_\alpha\right\rangle}{v_c}\frac{\left\langle c_{\alpha+\gamma}\right\rangle}{v_c}=\frac{1}{M\rho^2 v_c^2}\sum_{\alpha=1}^M\left\langle c_\alpha\right\rangle\frac{1}{N_\gamma}\sum_{|\gamma|=r}\left\langle c_{\alpha+\gamma}\right\rangle
\label{eq-93}
\ee
and
\be
\widetilde{h}(r)\approx\frac{1}{M\rho^2 v_c^2}\sum_{\alpha=1}^M\frac{1}{N_\gamma}\sum_{|\gamma|=r}\left\langle c_\alpha c_{\alpha+\gamma}\right\rangle\ln\frac{\left\langle c_\alpha c_{\alpha+\gamma}\right\rangle}{\left\langle c_\alpha\right\rangle\left\langle c_{\alpha+\gamma}\right\rangle}\,.
\label{eq-94}
\ee
While $\widetilde{g}(r)$ is the statistical average of an estimator whose histogram can be updated in the course of the simulation (see Eq.~(\ref{eq-86})), $\widetilde{g}_0(r)$ can only be estimated at the end of simulation, once $\langle c_\alpha\rangle$ has been evaluated for every $\alpha$ with an effort comparable to that made for the one-body entropy. Much more costly is the calculation of $\widetilde{h}(r)$, which should also be performed at the end of simulation after evaluating $\langle c_\alpha c_\beta\rangle$ for every $\alpha$ and $\beta$.

Using translational lattice symmetry, the radial distribution functions and $\widetilde{h}(r)$ of a crystal can also be written as:
\ba
\rho\widetilde{g}(r)&=&\int_{\cal C}{\rm d}^3r_1\int\frac{{\rm d}^2\Omega}{4\pi}\,\rho^{(2)}({\bf r}_1,{\bf r}_1+{\bf r})\,;\,\,\,\,\,\,\rho\widetilde{g}_0(r)=\int_{\cal C}{\rm d}^3r_1\int\frac{{\rm d}^2\Omega}{4\pi}\,\rho^{(1)}({\bf r}_1)\rho^{(1)}({\bf r}_1+{\bf r})\,;
\nonumber \\
\rho\widetilde{h}(r)&=&\int_{\cal C}{\rm d}^3r_1\int\frac{{\rm d}^2\Omega}{4\pi}\,\rho^{(2)}({\bf r}_1,{\bf r}_1+{\bf r})\ln\frac{\rho^{(2)}({\bf r}_1,{\bf r}_1+{\bf r})}{\rho^{(1)}({\bf r}_1)\rho^{(1)}({\bf r}_1+{\bf r})}\,,
\label{eq-95}
\ea
leading to simplifying Eqs.~(\ref{eq-85}), (\ref{eq-93}), and (\ref{eq-94}) into
\ba
\widetilde{g}(r)&=&\left\langle\sum_{\alpha=1}^{M/N}\delta_{c_\alpha,1}\frac{\sum_{|\gamma|=r}c_{\alpha+\gamma}}{4\pi r^2\Delta r\rho}\right\rangle\,;\,\,\,\,\,\,\widetilde{g}_0(r)=\sum_{\alpha=1}^{M/N}\left\langle c_\alpha\right\rangle\frac{\sum_{|\gamma|=r}\left\langle c_{\alpha+\gamma}\right\rangle}{4\pi r^2\Delta r\rho}\,;
\nonumber \\
\widetilde{h}(r)&=&\sum_{\alpha=1}^{M/N}\frac{1}{4\pi r^2\Delta r\rho}\sum_{|\gamma|=r}\left\langle c_\alpha c_{\alpha+\gamma}\right\rangle\ln\frac{\left\langle c_\alpha c_{\alpha+\gamma}\right\rangle}{\left\langle c_\alpha\right\rangle\left\langle c_{\alpha+\gamma}\right\rangle}\,.
\label{eq-96}
\ea
In the above formulae, the $\alpha$ index only runs over the cubes contained in a Wigner-Seitz/Voronoi cell of the lattice, while the $\beta$ sum is still carried out over all cubes in the simulation box.

\subsection{Numerical tests}

We first examine the shape of the structure functions $\widetilde{g}(r)$ and $\widetilde{g}_0(r)$ for hard spheres, choosing a $r$ resolution of $\Delta r=0.05$ (in units of the particle diameter $\sigma$). We take a system of $N=4000$ particles arranged in a fcc lattice with packing fraction $\eta=0.600$ (recall that the melting value is approximately 0.545). Periodic conditions are applied at the system boundary. In order to constrain the crystal in space, we keep one particle fixed during the simulation. As for $\widetilde{g}_0(r)$, we employ the Tarazona ansatz for $\alpha=95$ (see Eq.~(\ref{eq-39})), a value providing the best fit to the one-body density drawn from simulation.

We use the standard Metropolis Monte Carlo (MC) algorithm, constantly adjusting the maximum shift of a particle during equilibration until the fraction of accepted moves becomes close to $50\%$ (then, the maximum shift is no longer changed). We produce 50000 MC cycles in the equilibration run, whereas CE averages are computed over a total of further $2\times 10^5$ cycles. Our results are plotted in Figure 1. While at short distances $\widetilde{g}(r)$ and $\widetilde{g}_0(r)$ are rather different, as $r$ increases the oscillations of the two functions become closer and closer in amplitude.

\begin{figure}
\centering
\includegraphics[width=12 cm]{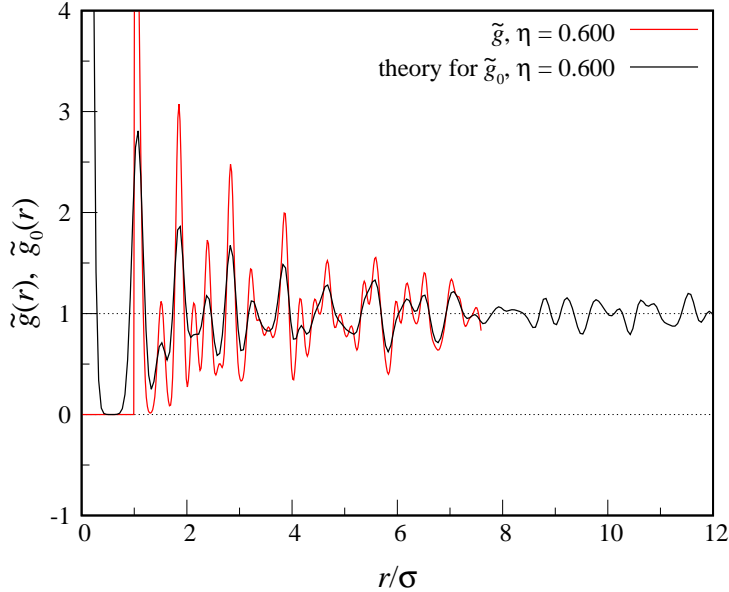}
\caption{We show a comparison between $\widetilde{g}(r)$ for a fcc crystal of hard spheres ($\eta=0.600$) and the $\widetilde{g}_0(r)$ function given in Eq.~(\ref{eq-39}), where the value of $\alpha$ (95) has been chosen such that the Tarazona ansatz (\ref{eq-21}) fits at best the one-body density drawn from simulation.}
\end{figure}

To obtain the one-body density with sufficient accuracy, we use a grid of about 50 points along each space direction in the unit cell. However, this grid resolution is too high for allowing the computation of $\widetilde{h}(r)$, as the memory requirements for processing the $\left\langle c_\alpha c_\beta\right\rangle$ data are very huge. On the other hand, a coarser grid is incompatible with the $\Delta r$ chosen.

To get closer to achieving our goal, i.e., to ascertain the $N$ dependence of the two-body entropy for a crystal, we consider a two-dimensional system --- hard disks. For this system the transformation from fluid to solid occurs in two stages, via an intermediate hexatic fluid phase~\cite{Bernard} (the transition from isotropic to hexatic fluid is first-order, whereas the hexatic-solid transition is continuous and occurs at $\eta=0.700$). We consider a system of $N=1152$ hard disks, arranged in a triangular crystal with packing fraction $\eta=0.800$, and a mesh consisting of about 80 points along each direction in the unit cell. Even though translational correlations are only quasi-long-ranged in an infinite two-dimensional crystal, when one of the particles is kept artificially fixed this specificity is lost and the (finite) two-dimensional crystal is made fully similar to a three-dimensional crystal. Also observe that an infinite two-dimensional crystal shares at least the same breaking of rotational symmetry typical of an infinite three-dimensional crystal.

As before, we first look at the structure functions drawn from simulation, $\widetilde{g}(r)$ and $\widetilde{g}_0(r)$. Our results are plotted in Figure 2, together with the $\widetilde{g}_0(r)$ function of Eq.~(\ref{eq-40}) for $\alpha=75$. For this $\alpha$ the matching between the two $\widetilde{g}_0$ functions is nearly perfect, indicating that the peaks of the one-body density are (to a high level of accuracy) Gaussian in shape. For $\eta=0.800$ we find $S_1/N=-2.156$.

\begin{figure}
\centering
\includegraphics[width=12 cm]{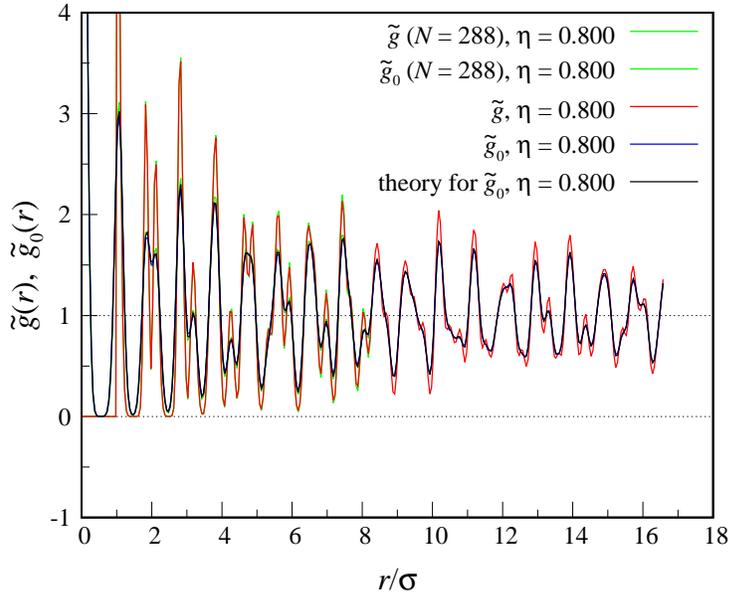}
\caption{Structure functions $\widetilde{g}(r)$ and $\widetilde{g}_0(r)$ for a triangular crystal of hard disks ($\eta=0.800$). We report data for two sizes, $N=288$ and $N=1152$. For comparison, we also plot the $\widetilde{g}_0(r)$ function in Eq.~(\ref{eq-40}) for $\alpha=75$. As is clear, the Tarazona ansatz represents an excellent model for the one-body density of the weakly-constrained hard-disk crystal.}
\end{figure}

In Figure 3, we show our main result, $\widetilde{h}(r)$, for $\eta=0.800$ and two different crystal sizes, $N=288$ and 1152. We point out that, in order to obtain these data, we had to run a separate simulation for each $r$, as the memory usage is rather extreme. To be sure, we have computed the $\widetilde{g}_0$ values in an independent way, i.e., using the same program loop written for $\widetilde{h}(r)$, eventually finding the same results as in Figure 2. Looking at Figure 3, we see that $\widetilde{h}(r)$ shows a series of peaks at neighbor positions and in the valleys within, taking preferentially positive values (meaning that its oscillations are not centered around zero). However, the damping of large-distance oscillations is too gradual to allow us assessing the nature of the asymptotic decay of $\widetilde{h}(r)$ and then compute $S_2$. We attempt a few explanations for this behavior of $\widetilde{h}(r)$: On one hand, the decay of $\widetilde{h}(r)$ may really be slow (at least in two dimensions), but $S_2$ would nonetheless be extensive, which implies a large $S_2/N$ value. It may as well be that constraining the crystal in space by hinging the position of one particle has a strong effect on the speed of $\widetilde{h}$ decay, which only a finite-size scaling of data can relieve. Indeed, when going from $N=288$ to $N=1152$ the values of $\widetilde{h}$ are slightly shifted downwards.

\begin{figure}
\centering
\includegraphics[width=12 cm]{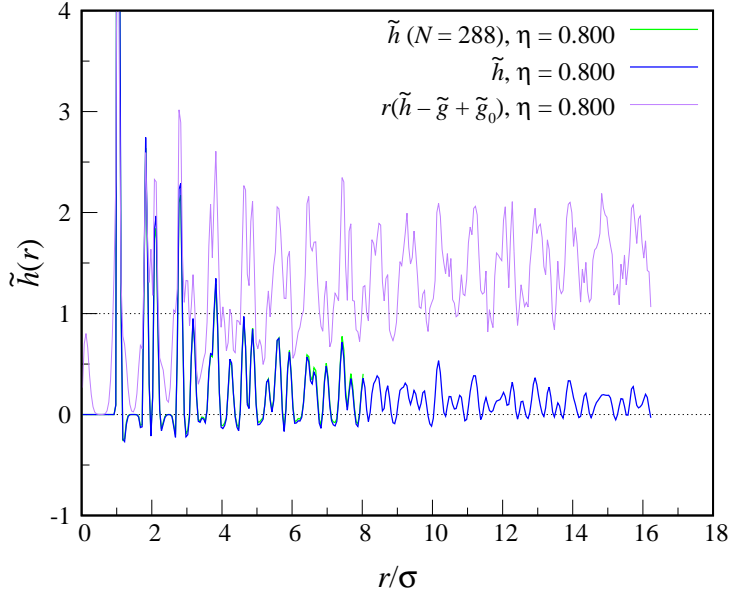}
\caption{The function $\widetilde{h}(r)$ for hard disks ($\eta=0.800$). As in Figure 2, data for two sizes are shown, namely $N=288$ and $N=1152$. It appears that the oscillations of $\widetilde{h}(r)$ decay very slowly, which implies slow convergence of the integrand in Eq.~(\ref{eq-36}) to zero.}
\end{figure}

In summary, we have not reached any clear demonstration of $S_2$ extensivity in a crystal. This task has proved to be very hard to settle numerically. Our hope is that, based on our preparatory work, other Authors with more powerful computational resources at their disposal can push the numerical analysis forward and eventually come up with a definite solution of the problem.

\section{Conclusions}

In this paper, we inquire into the possibility of extending the zero-RMPE criterion, a popular one-phase criterion of freezing for simple fluids, to also cover the melting of a solid. After revisiting the derivation of the entropy MPCE in the canonical ensemble, we argue that the formula applies for a crystal too. We exploit lattice symmetries to constrain the structure of one- and two-body densities, so as to gain as much information as possible on the first few terms in the entropy expansion. While this is enough to prove that the crystal one-body entropy is an extensive quantity, the information obtained is not sufficient to hold the same for the two-body entropy, whose scaling with the size of the crystal remains elusive. We have thus attempted to clarify the question numerically, but we have been faced against an insurmountable obstacle in computational and memory limitations. To alleviate the problem, we have turned towards a two-dimensional case, namely hard disks, but with poor results: the structure function that must be integrated over distances to obtain the two-body entropy is weakly convergent to zero. In the near future, we plan to check whether the situation is more favorable for a different two-dimensional interaction, either endowed with an attractive tail (e.g., the Lennard-Jones potential) or provided with a soft core (for example, a Gaussian repulsion).

\acknowledgments{This work has benefited from computer facilities made available by the PO-FESR 2007-2013 Project MedNETNA (Mediterranean Network for Emerging Nanomaterials).}

\appendix
\section{Truncating the entropy expansion}

We hereafter give an interpretation of the successive estimates of $S_N^{\rm exc}/k_B=-\langle\ln P_{1\ldots N}\rangle$ obtained by stopping the expansion (\ref{eq-13}) at a given order of correlations. We show that each truncated entropy expansion can be arranged in the form $-\langle\ln P^\star_{1\ldots N}\rangle$, where $P^\star_{1\ldots N}$ is a functional of all the MDFs up to $n$-th order, for $n=1,2,\ldots,N$ (however, without claiming that $P^\star_{1\ldots N}$ represents a proper, i.e., normalized distribution). Our method resembles the one originally devised by H. S. Green to express the canonical entropy of a $N$-particle fluid in terms of correlation functions~\cite{Green}. While H. S. Green correctly inferred the first three terms in the expansion, he did not provide a general recipe to obtain the further terms recursively.

For $N=1$ there is only one MDF, $P_1\equiv P^{(1)}({\bf r}_1)$, in terms of which a fully symmetric approximation to $P_{1\ldots N}$ can be constructed:
\be
{\rm 1st-order\,\,approximation}:\qquad P^\star_{1\ldots N}=\prod_i^NP_i\,.
\label{a-1}
\ee
Notice that $-\langle\ln P^\star_{1\ldots N}\rangle=S_N^{(1)}/k_B$.

To obtain a better approximation we consider a system of two particles. Since
\be
P_{12}=P_1P_2\times\frac{P_{12}}{P_1P_2}\,,
\label{a-2}
\ee
we see that $P_{12}$ is the product of the 1st-order approximation (\ref{a-1}) times a correction factor $P_{12}/(P_1P_2)=Q_{12}$. Assuming that in a $N$-particle system each distinct pair of particles contributes the same factor to $P^\star_{1\ldots N}$, we arrive at the
\be
{\rm 2nd-order\,\,approximation}:\qquad P^\star_{1\ldots N}=\prod_i^NP_i\prod_{i<j}^NQ_{ij}\,.
\label{a-3}
\ee
The number of factors in the second product is $N(N-1)/2$. For this $P^\star_{1\ldots N}$ we obtain
\be
-\langle\ln P^\star_{1\ldots N}\rangle=-N\int P_1\ln P_1-{N\choose 2}\int P_{12}\ln Q_{12}=S_N^{(2)}/k_B\,.
\label{a-4}
\ee

Moving to $N=3$, we observe that
\be
P_{123}=P_1P_2P_3Q_{12}Q_{13}Q_{23}\times\frac{Q_{123}}{Q_{12}Q_{13}Q_{23}}\,,
\label{a-5}
\ee
which is the second-order approximation to $P^\star_{123}$ times a correction factor. In the event that each distinct triplet of particles contributes the same factor to $P^\star_{1\ldots N}$, we obtain the
\be
{\rm 3rd-order\,\,approximation}:\qquad P^\star_{1\ldots N}=\prod_i^NP_i\prod_{i<j}^NQ_{ij}\prod_{i<j<k}^N\frac{Q_{ijk}}{Q_{ij}Q_{ik}Q_{jk}}\,.
\label{a-6}
\ee
Notice that a different expression for the latter ratio is
\be
\frac{Q_{ijk}}{Q_{ij}Q_{ik}Q_{jk}}=\frac{P_{ijk}}{\frac{P_{ij}P_{ik}P_{jk}}{P_iP_jP_k}}\,.
\label{a-7}
\ee
The number of factors in the third product is $N(N-1)(N-2)/6$. For this $P^\star_{1\ldots N}$, the approximate entropy is
\small
\be
-\langle\ln P^\star_{1\ldots N}\rangle=-N\int P_1\ln P_1-{N\choose 2}\int P_{12}\ln Q_{12}-{N\choose 3}\left[\int P_{123}\ln Q_{123}-{3\choose 2}\int P_{12}\ln Q_{12}\right]=\frac{S_N^{(3)}}{k_B}\,.
\label{a-8}
\ee
\normalsize

We can similarly proceed to derive higher-order approximations. The 4-body MDF of a system of $N=4$ particles is trivially decomposed as
\ba
P_{1234}&=&P_1P_2P_3P_4Q_{12}Q_{13}Q_{14}Q_{23}Q_{24}Q_{34}\frac{Q_{123}}{Q_{12}Q_{13}Q_{23}}\frac{Q_{124}}{Q_{12}Q_{14}Q_{24}}\frac{Q_{134}}{Q_{13}Q_{14}Q_{34}}\frac{Q_{234}}{Q_{23}Q_{24}Q_{34}}
\nonumber \\
&\times&\frac{Q_{1234}}{Q_{12}Q_{13}Q_{14}Q_{23}Q_{24}Q_{34}\frac{Q_{123}}{Q_{12}Q_{13}Q_{23}}\frac{Q_{124}}{Q_{12}Q_{14}Q_{24}}\frac{Q_{134}}{Q_{13}Q_{14}Q_{34}}\frac{Q_{234}}{Q_{23}Q_{24}Q_{34}}}
\nonumber \\
&=&P_1P_2P_3P_4Q_{12}Q_{13}Q_{14}Q_{23}Q_{24}Q_{34}\frac{Q_{123}}{Q_{12}Q_{13}Q_{23}}\frac{Q_{124}}{Q_{12}Q_{14}Q_{24}}\frac{Q_{134}}{Q_{13}Q_{14}Q_{34}}\frac{Q_{234}}{Q_{23}Q_{24}Q_{34}}
\nonumber \\
&\times&\frac{Q_{1234}}{\frac{Q_{123}Q_{124}Q_{134}Q_{234}}{Q_{12}Q_{13}Q_{14}Q_{23}Q_{24}Q_{34}}}\,,
\label{a-9}
\ea
whence the
\small
\be
{\rm 4th-order\,\,approximation}:\qquad P^\star_{1\ldots N}=\prod_i^NP_i\prod_{i<j}^NQ_{ij}\prod_{i<j<k}^N\frac{Q_{ijk}}{Q_{ij}Q_{ik}Q_{jk}}\prod_{i<j<k<l}^N\frac{Q_{ijkl}}{\frac{Q_{ijk}Q_{ijl}Q_{ikl}Q_{jkl}}{Q_{ij}Q_{ik}Q_{il}Q_{jk}Q_{jl}Q_{kl}}}\,.
\label{a-10}
\ee
\normalsize
Notice that a different expression for the latter ratio is
\be
\frac{Q_{ijkl}}{\frac{Q_{ijk}Q_{ijl}Q_{ikl}Q_{jkl}}{Q_{ij}Q_{ik}Q_{il}Q_{jk}Q_{jl}Q_{kl}}}=\frac{P_{ijkl}}{\frac{P_{ijk}P_{ijl}P_{ikl}P_{jkl}}{\frac{P_{ij}P_{ik}P_{il}P_{jk}P_{jl}P_{kl}}{P_iP_jP_kP_l}}}\,.
\label{a-11}
\ee
For this $P^\star_{1\ldots N}$ we obtain
\small
\ba
-\langle\ln P^\star_{1\ldots N}\rangle&=&-N\int P_1\ln P_1-{N\choose 2}\int P_{12}\ln Q_{12}-{N\choose 3}\left[\int P_{123}\ln Q_{123}-{3\choose 2}\int P_{12}\ln Q_{12}\right]
\nonumber \\
&-&{N\choose 4}\left[\int P_{1234}\ln Q_{1234}-{4\choose 3}\int P_{123}\ln Q_{123}+{4\choose 2}\int P_{12}\ln Q_{12}\right]=S_N^{(4)}/k_B\,.
\label{a-12}
\ea
\normalsize

Eventually, with the last $N$th-order approximation we recover the exact distribution, namely $P^\star_{1\ldots N}=P_{1\ldots N}$, and the full entropy. Notice that, except for $n=1$ and $N$, the $n$th-order functional $P^\star_{1\ldots N}$ is {\em not} normalized.

We now provide a formalization of the procedure sketched above. For each value of $n$ and each grouping $I_n=\{i_1,i_2,\ldots i_n\}$ of $n$ particle indices, we write $P(I_n)\equiv P_{i_1\ldots i_n}$ as a product of positive cumulant factors to be determined recursively, that is
\be
P(I_n)=\prod_{S_1\subset I_n}C(S_1)\cdots\prod_{S_{n-1}\subset I_n}C(S_{n-1})\times C(I_n)\,,
\label{a-13}
\ee
where $\prod_{S_k\subset I_n}$ indicates the product over all $k$-tuples of distinct entries from $I_n$ --- there are $n\choose k$ factors in the product $\prod_{S_k\subset I_n}$. As shown before, $C(\{i\})=P_i,C(\{i,j\})=P_{ij}/(P_iP_j),C(\{i,j,k\})=P_{ijk}P_iP_jP_k/(P_{ij}P_{ik}P_{jk})$, and so on. Taking the logarithm of (\ref{a-13}) we obtain:
\be
\ln P(I_n)=\ln C(I_n)+\sum_{S_{n-1}\subset I_n}\ln C(S_{n-1})+\ldots+\sum_{S_1\subset I_n}\ln C(S_1)\,,
\label{a-14}
\ee
which can be solved with respect to cumulants by the M\"obius inversion formula (see, e.g., Eqs.~3 and 4 of Ref.~\cite{An}):
\be
\ln C(I_n)=\ln P(I_n)-\sum_{S_{n-1}\subset I_n}\ln P(S_{n-1})+\ldots+(-1)^{n-1}\sum_{S_1\subset I_n}\ln P(S_1)\,,
\label{a-15}
\ee
which is in turn equivalent to writing
\be
C(I_n)=\frac{P(I_n)}{\frac{\prod_{S_{n-1}\subset I_n}P(S_{n-1})}{\frac{\vdots}{\prod_{S_1\subset I_n}P(S_1)}}}\,.
\label{a-16}
\ee
Equations (\ref{a-7}) and (\ref{a-11}) are just particular cases of the above formula, respectively for $n=3$ and $n=4$. Once the cumulants have been determined, the functional $P_{1\ldots N}^*$ of $M$-th order (for $M=1,\ldots,N$) can be written, by an obvious change of notation, as
\be
P_{1\ldots N}^*=\prod_i^NC_i\prod_{i<j}^NC_{ij}\cdots\prod_{i_1<\ldots<i_M}^NC_{i_1\ldots i_M}\,,
\label{a-17}
\ee
leading to
\be
-\left\langle\ln P_{1\ldots N}^*\right\rangle=-N\int P_1\ln C_1-{N\choose 2}\int P_{12}\ln C_{12}-\ldots-{N\choose M}\int P_{12\ldots M}\ln C_{12\ldots M}\,.
\label{a-18}
\ee
In view of Eq.~(\ref{a-16}), the above quantity is nothing but $S_N^{(M)}$.



\begin{thebibliography}{999}
\bibitem{Green} Green, H. S. {\em The Molecular Theory of Fluids}; North Holland: Amsterdam, The Netherlands, 1952; pp. 70--73.

\bibitem{Nettleton} Nettleton, R. E. and Green, M. S. Expression in Terms of Molecular Distribution Functions for the Entropy Density in an Infinite System. {\em J. Chem. Phys.} {\bf 1958}, {\em 29}, 1365--1370.

\bibitem{Baranyai} Baranyai, A. and Evans, D. J. Direct entropy calculation from computer simulation of liquids. {\em Phys. Rev. A} {\bf 1989}, {\em 40}, 3817--3822.

\bibitem{Schlijper} Schlijper, A. G. Convergence of the cluster-variation method in the thermodynamic limit. {\em Phys. Rev. B} {\bf 1983}, {\em 27}, 6841-6848.

\bibitem{An} An, G. A Note on the Cluster Variation Method. {\em J. Stat. Phys.} {\bf 1988}, {\em 52}, 727--734.

\bibitem{Pelizzola} Pelizzola, A. Cluster variation method in statistical physics and probabilistic graphical models. {\em J. Phys. A} {\bf 2005}, {\em 38}, R309-R339.

\bibitem{Hernando} Hernando, J. A. Thermodynamic potentials and distribution functions: I. A general expression for the entropy. {\em Mol. Phys.} {\bf 1990}, {\em 69}, 319-326.

\bibitem{Prestipino1} Prestipino, S. and Giaquinta, P. V. Statistical entropy of a lattice-gas model: multiparticle correlation expansion. {\em J. Stat. Phys.} {\bf 1999}, {\em 96}, 135--167; Erratum: {\em ibid.} {\bf 2000}, {\em 98}, 507--509.

\bibitem{Prestipino2} Prestipino, S. and Giaquinta, P. V. The entropy multiparticle-correlation expansion for a mixture of spherical and elongated particles. {\em J. Stat. Mech.: Theor. Exp.} {\bf 2004}, P09008.

\bibitem{D'Alessandro} D'Alessandro, M. Multiparticle correlation expansion of relative entropy in lattice systems. {\em J. Stat. Mech.: Theor. Exp.} {\bf 2016}, 073201.

\bibitem{Maffioli} Maffioli, L.; Clisby, N.; Frascoli, F.; and Todd, B. D. Computation of the equilibrium three-particle entropy for dense atomic fluids by molecular dynamics simulation. {\em J. Chem. Phys.} {\bf 2019}, {\em 151}, 164102.

\bibitem{Abramo} Abramo, M. C.; Caccamo, C.; Costa, D.; Giaquinta, P. V.; Malescio, G.; Muna\`o, G.; Prestipino, S. On the determination of phase boundaries via thermodynamic integration across coexistence regions. {\em J. Chem. Phys.} {\bf 2015}, {\em 142}, 214502.

\bibitem{Giaquinta1} Giaquinta, P. V. and Giunta G. About entropy and correlations in a fluid of hard spheres. {\em Physica A} {\bf 1992}, {\em 187}, 145--158.

\bibitem{Giaquinta2} Giaquinta, P. V.; Giunta, G.; and Prestipino Giarritta, S. Entropy and the freezing of simple liquids. {\em Phys. Rev. A} {\bf 1992}, {\em 45}, R6966--R6968.

\bibitem{Saija1} Saija, F.;  Pastore, G.;  Giaquinta, P. V. Entropy and Fluid-Fluid Separation in Nonadditive Hard-Sphere Mixtures. {\em J. Phys. Chem. B} {\bf 1998}, {\em 102}, 10368--10371.

\bibitem{Donato} Donato, M. G.; Prestipino, S.; and Giaquinta, P. V. Entropy and multi-particle correlations in two-dimensional lattice gases. {\em Eur. Phys. J. B} {\bf 1999}, {\em 11}, 621--627.

\bibitem{Saija2} Saija, F.; Prestipino, S.; and Giaquinta, P. V. Entropy, correlations, and ordering in two dimensions. {\em J. Chem. Phys.} {\bf 2000}, {\em 113}, 2806--2813.

\bibitem{Costa} Costa, D.; Micali, F.; Saija, F.; Giaquinta, P. V. Entropy and Correlations in a Fluid of Hard Spherocylinders: The Onset of Nematic and Smectic Order. {\em J. Phys. Chem. B} {\bf 2002}, {\em 106}, 12297--12306.

\bibitem{Prestipino3} Prestipino, S. Analog of surface preroughening in a two-dimensional lattice Coulomb gas. {\em Phys. Rev. E} {\bf 2002}, {\em 66}, 021602.

\bibitem{Saija3} Saija, F.; Saitta, A. M.; Giaquinta, P. V. Statistical entropy and density maximum anomaly in liquid water. {\em J. Chem. Phys.} {\bf 2003}, {\em 119}, 3587--3589.

\bibitem{Speranza} Speranza, C.; Prestipino, S.; Malescio, G.; and Giaquinta, P. V. Phase behavior of a fluid with a double Gaussian potential displaying waterlike features. {\em Phys. Rev. E} {\bf 2014}, {\em 90}, 012305.

\bibitem{Prestipino4} Prestipino, S. and Malescio, G. Characterization of the structural collapse undergone by an unstable system of ultrasoft particles. {\em Physica A} {\bf 2016}, {\em 457}, 492--505.

\bibitem{Banerjee} Banerjee, A.; Nandi, M. K.; Sastry, S.; Bhattacharyya, S. M. Determination of onset temperature from the entropy for fragile to strong liquids. {\em J. Chem. Phys.} {\bf 2017}, {\em 147}, 024504.

\bibitem{Santos} Santos, A.; Saija, F.; and Giaquinta, P. V. Residual Multiparticle Entropy for a Fractal Fluid of Hard Spheres. {\em Entropy} {\bf 2018}, {\em 20}, 544.

\bibitem{Frenkel} Frenkel, D. Order through entropy. {\em Nature Mater.} {\bf 2015}, {\em 14}, 9--12.

\bibitem{Speedy} Speedy, R. J. The entropy of a glass. {\em Mol. Phys.} {\bf 1993}, {\em 80}, 1105--1120.

\bibitem{Berthier} Berthier, L.; Ozawa, M.; and Scalliet, C. Configurational entropy of glass-forming liquids. {\em J. Chem. Phys.} {\bf 2019}, {\em 150}, 160902.

\bibitem{Baus} See, e.g., Baus, M. and Tejero, C. F. {\em Equilibrium Statistical Physics}; Springer: Berlin, Germany, 2008; pp. 61--63.

\bibitem{Tarazona} Tarazona, P. A density functional theory of melting. {\em Mol. Phys.} {\bf 1984}, {\em 52}, 81--96.

\bibitem{Prestipino5} Prestipino, S. and Giaquinta, P. V. Ground state of weakly repulsive soft-core bosons on a sphere. {\em Phys. Rev. A} {\bf 2019}, {\em 99}, 063619.

\bibitem{Prestipino6} Prestipino, S.; Sergi, A.; and Bruno E. Freezing of soft-core bosons at zero temperature: a variational theory. {\em Phys. Rev. B} {\bf 2018}, {\em 98}, 104104.

\bibitem{Rascon} Rasc\'on, C.; Mederos, L.; and Navascu\'es, G. Thermodynamic consistency of the hard-sphere solid distribution function. {\em J. Chem. Phys.} {\bf 1996}, {\em 105}, 10527--10534.

\bibitem{Prestipino7} Prestipino Giarritta, S.; Ferrario, M.; and Giaquinta, P. V. Statistical geometry of hard particles on a sphere: analysis of defects at high density. {\em Physica A} {\bf 1993}, {\em 201}, 649--665.

\bibitem{Hansen} Hansen, J.-P. and McDonald, I. R. {\em Theory of Simple Liquids}; Academic: Oxford, United Kingdom, 2013.

\bibitem{Gernoth} Gernoth, K. A. Spatial Microstructure of Quantum Crystals. {\em J. Low. Temp. Phys.} {\bf 2002}, {\em 126}, 725--730.

\bibitem{Bernard} Bernard, E. P. and Krauth, W. Two-Step Melting in Two Dimensions: First-Order Liquid-Hexatic Transition. {\em Phys. Rev. Lett.} {\bf 2011}, {\em 107}, 155704. 
\end{thebibliography}
\end{document}